\documentstyle[11pt,psfig]{article}

\makeatletter
\@addtoreset{equation}{section}
\makeatother

\relax

\def\4R{{{}^{(4)}R}} 

\def\K5{{\kappa}}
\def\K52{{\kappa^2}}

\newcommand{\ii}{i}
\newcommand{\jj}{j}

\newcommand{\da}{\dot{a}}
\newcommand{\db}{\dot{b}}
\newcommand{\dn}{\dot{n}}
\newcommand{\dda}{\ddot{a}}
\newcommand{\ddb}{\ddot{b}}

\newcommand{\pa}{a^{\prime}}
\newcommand{\pb}{b^{\prime}}
\newcommand{\pn}{n^{\prime}}
\newcommand{\ppa}{a^{\prime \prime}}

\newcommand{\ppn}{n^{\prime \prime}}
\newcommand{\fda}{\frac{\da}{a}}
\newcommand{\fdb}{\frac{\db}{b}}
\newcommand{\fdn}{\frac{\dn}{n}}
\newcommand{\fdda}{\frac{\dda}{a}}
\newcommand{\fddb}{\frac{\ddb}{b}}

\newcommand{\fpa}{\frac{\pa}{a}}
\newcommand{\fpb}{\frac{\pb}{b}}
\newcommand{\fpn}{\frac{\pn}{n}}
\newcommand{\fppa}{\frac{\ppa}{a}}

\newcommand{\fppn}{\frac{\ppn}{n}}

\def\be{\begin{equation}}
\def\ee{\end{equation}}
\def\bs{\begin{subequations}}
\def\es{\end{subequations}}
\def\g{\gamma}
\def\vp{\varphi}

\def\l{\lambda}

\def\d{\partial}

\def\r{\rho}

\def\k{\kappa}
\def\m{\mu}
\def\n{\nu}
\def\r{\rho}
\def\s{\sigma}
\def\t{\tau}
\def\a{\alpha}
\def\b{\beta}

\newcommand{\fig}[1]{Fig.~(\ref{#1})}

\newcommand{\GeV}{\mbox{GeV}}
\newcommand{\cm}{\mbox{cm}}
\newcommand\fverb{\setbox\pippobox=\hbox\bgroup\verb}
\newcommand\fverbdo{\egroup\medskip\noindent%
                        \fbox{\unhbox\pippobox}\ }
\newcommand\fverbit{\egroup\item[\fbox{\unhbox\pippobox}]}
\newbox\pippobox

\begin{document}
\begin{flushright}
\hfill{BROWN-HET-1209}\\
\hfill{hep-th/0003086}\\
\hfill{October 2000}\\
\end{flushright}
\vspace{24pt}
\newcommand{\tamphys}{\it Brown University, Department of Physics, 
\\ Providence, RI 02912, USA}

\newcommand{\auth}{Damien A. Easson\footnote{easson@het.brown.edu}}

\begin{center}
{ \large
{\bf THE INTERFACE OF COSMOLOGY \\ WITH STRING AND M(ILLENNIUM) THEORY}}

\vspace{36pt}

\auth

\vspace{10pt}

{\tamphys}

\vspace{44pt}

\underline{ABSTRACT}

\end{center}

The purpose of this review is to discuss recent developments occurring at the
interface of cosmology with string and M-theory.  We begin with a short
review of 1980s string cosmology and the Brandenberger-Vafa mechanism for explaining
spacetime dimensionality.  It is shown how this scenario has been modified to include
the effects of
p-brane gases in the early universe.  We then introduce the Pre-Big-Bang scenario (PBB), 
Ho\v{r}ava-Witten heterotic M-theory and the work of Lukas, Ovrut and Waldram, 
and end with a discussion of large extra dimensions, the Randall-Sundrum model and
Brane World cosmologies.

\vspace{20pt}


{\vfill\leftline{}\vfill}
\pagebreak
\setcounter{page}{1}

\tableofcontents
\addtocontents{toc}{\protect\setcounter{tocdepth}{2}}
\newpage




\section{Introduction}

 In recent years there have been many exciting advances in our understanding of M-theory -- our
best candidate for the fundamental theory of everything.  The theory claims to describe physics
appropriately in regions of space with high 
energies and large curvature scales.  As these characteristics are exactly those found
in the initial conditions of the universe it is
only natural to incorporate M-theory into models of early universe cosmology. 

The necessity
to search for alternatives to the Standard Big-Bang (SBB) 
model of cosmology stems from a number of 
detrimental problems such as
the horizon, flatness, structure formation and cosmological constant problems.
Although inflationary models have managed to address many 
of these issues, inflation, at least in its current formulation,
does not explain everything.  In particular, inflation fails to address 
the fluctuation, super-Planck scale physics, initial singularity
and cosmological constant problems as discussed in \cite{ref:brandenberger1999}.

At the initial singularity, physical invariants
such as the Ricci scalar, $R$, blow up.  Other 
measurable quantities, for example
temperature and energy density also become infinite.  
From the Hawking-Penrose singularity theorems we know that such spacetimes are
geodesically incomplete.  So, when we ask the question of how the universe began, the
inevitable and unsatisfactory answer is that we don't know.  The physics required to
understand this epoch of the early universe
is necessarily rooted in a theory of quantum gravity.  Presently, string theory is the only candidate
for such a unifying theory.  It is therefore logical to study the ways in
which it changes our picture of cosmology.  Although an ambitious aspiration,
we hope that M-theory will solve the above
mentioned dilemmas and provide us with a complete description of the evolution of
the universe.

In this analysis, we must proceed with caution.  
Our present understanding of M-theory is extremely limited, as is our understanding of
cosmology before the first $10^{-43}$ seconds.  Nevertheless, it is clear that the study of string cosmology 
is essential to the development of string theory, and extremely important for our 
understanding of the early universe.
  
The purpose of this article is to introduce some of the most promising work
and themes under investigation in string cosmology.  We begin with
a brief, qualitative introduction to M-theory in Section \ref{mtheory}.  

In Section \ref{dimen} we review the work of Brandenberger and Vafa 
\cite{ref:brandenberger1989} in which the 1980s version of string theory is used to
solve the initial singularity problem and
in an attempt to explain why we live in four macroscopic dimensions despite the
fact that string
theory seems to predict the wrong number of dimensions, namely ten.  We then explain how
this scenario has been updated in order to include the effects of $p$-branes \cite{ref:alexanderet}.

Section \ref{PBB} provides a brief introduction to the Pre-Big-Bang scenario 
\cite{ref:veneziano1991}-\cite{ref:lidseyet1999}.
This is a theory based on the low energy effective action
for string theory, developed in the early 1990s by Gasperini and Veneziano.  

Another
promising attempt to combine M-theory with cosmology, that of Lukas, Ovrut and
Waldram \cite{ref:lukaset1998}, is presented in Section \ref{heterotic}.  
Their work is based on the model of heterotic M-theory constructed by Ho\v{r}ava and Witten 
and is inspired
by eleven dimensional supergravity, the low energy limit of M-theory.  The motivation
for this work was to construct a toy cosmological model from the most fundamental theory we know.

The final section (\ref{led}) 
reviews some models involving large extra dimensions.  This section begins with a short
introduction to the hierarchy problem of standard model particle physics and explains how it may be solved
using large extra dimensions.  ``Brane World" scenarios are then discussed focusing primarily on
the models of Randall and Sundrum \cite{ref:randallet1999, ref:randallsu1999}, 
where our four dimensional universe emerges as the world volume of a three brane.  
The cosmologies of such theories are reviewed, and we briefly comment on their incorporation into
supergravity models, string theory and the AdS/CFT correspondence.

The sections in this review are
presented more or less chronologically.\footnote{This review is in no way comprehensive.  
As it is impossible to discuss all aspects of string cosmology I have included
a large list of references at the end.  Some of the topics I will not
cover in the text may be found there.  For discussions of p-brane dynamics and cosmology
see \cite{ref:parket1999}-\cite{ref:luet1999}, \cite{ref:alexanderet}.
For recent reviews on other cosmological aspects of M-theory see    
\cite{ref:dine2000, ref:greene1999, ref:banks1999}.  For some ideas on 
radically new cosmologies from M-theory see e.g. 
\cite{ref:lizziet2000}-\cite{ref:verlinde1999}.}

\section{M-Theory} \label{mtheory}

For several years now, we have known that there are five consistent formulations
of superstring theory.  The five theories are ten-dimensional, two having $N = 2$ 
supersymmetry known as Type IIA and Type IIB and three having $N=1$ supersymmetry,
Type I, $SO(32)$ heterotic and $E_8 \times E_8$ heterotic.  Recently, duality symmetries
between the various theories have been discovered, leading to the conjecture that
they all represent different corners of a large, multidimensional moduli space of a unified
theory named, M-theory.  Using dualities 
we have discovered that there is a sixth branch to the M-theory moduli 
space (see \fig{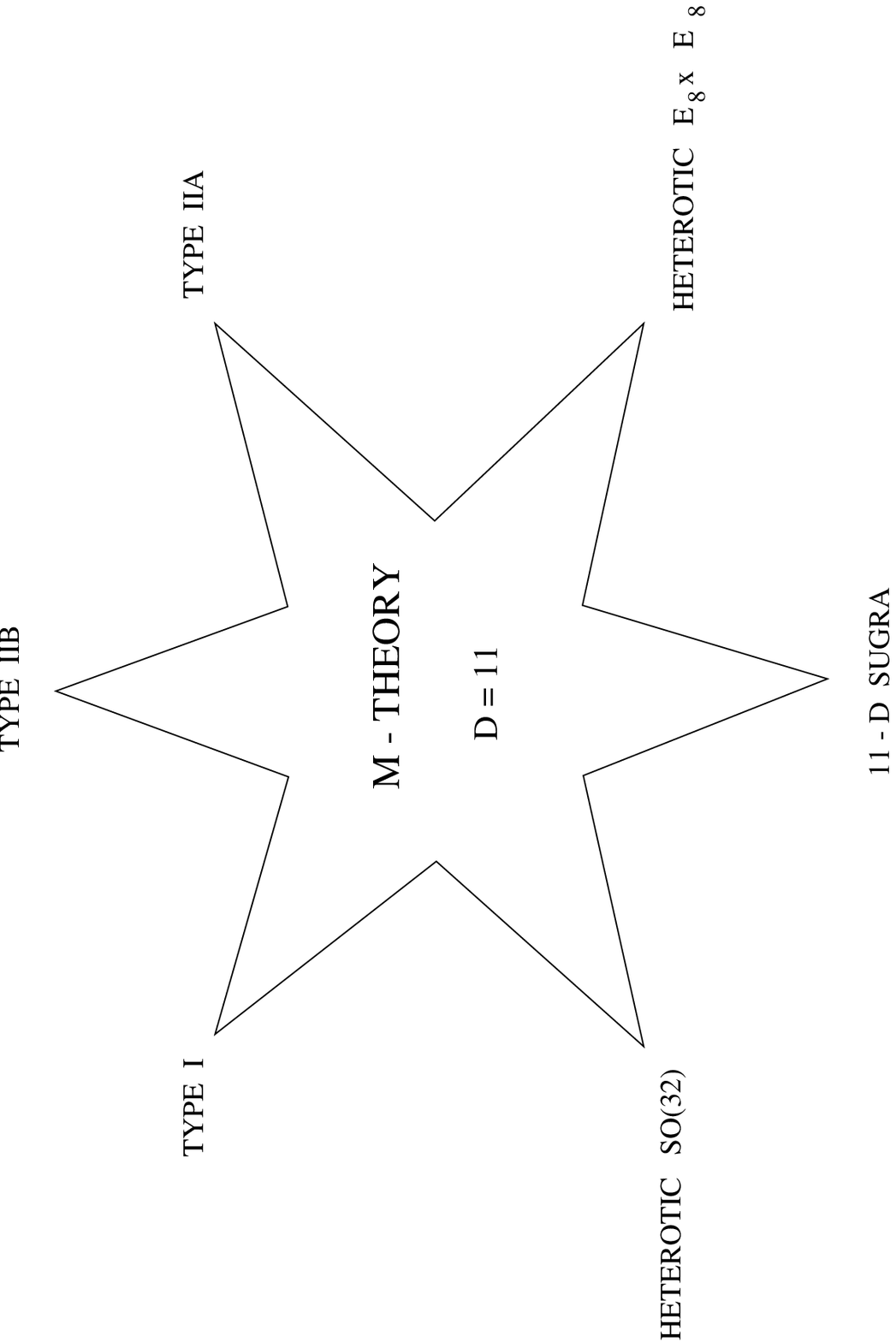}) corresponding to eleven-dimensional 
supergravity \cite{ref:witten1995}.    

\vspace{.75cm}
\hglue 1.5cm
\psfig{figure=moduli.eps,height=8cm,angle=-90}\label{moduli.eps}
\begin{quote}
\scriptsize Figure \ref{moduli.eps}: This is a slice of the eleven-dimensional
moduli space of M-theory.  Depicted are the five ten-dimensional string 
theories and eleven-dimensional supergravity, which is identified with
the low energy limit of M-theory.
\end{quote}

It is possible that using these
six cusps of the moduli space we have already identified the fundamental degrees
of freedom of the entire nonperturbative M-theory, but that their full significance
has yet to be appreciated.  A complete understanding and consistent formulation
of M-theory is the ultimate challenge for string theorists today and will take
physicists into the new M(illennium).

\section{Superstrings and Spacetime Dimensionality} \label{dimen}

Perhaps the greatest embarrassment of string theory is the dimensionality problem.
We perceive our universe to be four dimensional, yet string theory seems
to naively predict the wrong number of dimensions, namely ten.  The typical 
resolution to this apparent conflict is to say that six of the dimensions are
curled up on a Planckian sized manifold.  The following question naturally arises, why
is there a six/four dimensional split between the small/large dimensions?  Why not
four/six, or seven/three?
Although there is still no official answer to this question, a possible explanation emerges from
cosmology and the work of Brandenberger and Vafa \cite{ref:brandenberger1989}
which we will summarize in this section.  We will then show how it is possible
to generalize this scenario of the 1980s to incorporate our current understanding of 
string theory \cite{ref:alexanderet}.

\subsection{Duality} \label{duality}

Before diving into the specifics of the BV model we review some basics of string dualities and thermodynamics.
Consider the dynamics of strings moving in a nine-dimensional box with sides
of length $R$.  We impose periodic boundary conditions for both bosonic and fermionic
degrees of freedom, so we are effectively considering
string propagation in a torus.  What types of objects are in our box?  For one, there
are oscillatory modes corresponding to vibrating stationary strings.  Then, there
are momentum modes which are strings moving in the box with fourier mode $n$ and 
momentum
\be\label{eq:moment}
p = n/R
\,.
\ee
There are also winding modes which are strings that stretch across the box (wrapped around the torus)
with energy given by
\be\label{eq:winding}
\omega = mR
\,,
\ee
where $m$ is the number of times the string winds around the torus.

We now make the remarkable observation, that the spectrum of this system remains unchanged
under the substitution
\be\label{eq:tdual}
R \rightarrow \frac{1}{R}
\,,
\ee
(provided we switch the roles of $m$ and $n$).  This symmetry is known as T-duality
\cite{ref:polchinski1998} and
is a symmetry of the entire M-theory, not just the spectrum of this particular
model.  T-duality leads
us to the startling conclusion that any physical process in a box of radius $R$ is
equivalent to a dual physical process in a box of radius $1/R$.  In other words,
one can show that scattering amplitudes for dual processes are equal.  Hence, we 
have discovered that distance, which is an invariant concept in general relativity (GR), is 
\it{not} \rm an invariant concept in string theory.  In fact, we will see that
many invariant notions in GR are not invariant notions in string theory.  These deviations
from GR are especially noticeable for small distance scales where the Fourier modes
of strings become heavier (\ref{eq:moment}) and less energetically favorable,
while the winding modes become light (\ref{eq:winding})
and are therefore more easy to create.

\subsection{Thermodynamics of Strings} \label{thermo}

Before discussing applications of t-duality to cosmology let
us review a few useful calculations of string thermodynamics.
The primary assumption we will make for the following discussion
is that the string coupling is sufficiently small
so that we may ignore the gravitational back reaction of thermodynamical string
condensates on the spacetime geometry.
 
String thermodynamics predicts the existence of a maximum temperature
known as the Hagedorn temperature ($T_H$)
above which the canonical ensemble approach
to thermodynamics breaks down \cite{ref:hagedorn1965}.  
This is due to the divergence of the partition function
because of string states which exponentially increase as
\be\label{eq:states}
 d(E) \propto E^{-p} \exp{(\b_H E)}
 \,,
\ee
where $p > 0$.  The partition function is easily calculated,
\be\label{eq:partition}
 Z = \sum_i \exp(-\b E_i)
\,,
\ee
which diverges for $\b < \b_H$, or $T>T_H$.\footnote{For more on string
thermodynamics see e.g. \cite{ref:hagedorn1965}-\cite{ref:meanaet1999}.}

\subsection{The BV Mechanism and the Early Universe} \label{early}

Consider the following toy model of a superstring-filled early universe.
Besides the assumption of small coupling stated in section \ref{thermo},
we also assume that the evolution of the universe is adiabatic and
make some assumptions about the size and shape of the universe.  

Before the
work of Brandenberger and Vafa, it was typical to speak about the process of ``spontaneous 
compactification" of six of the ten dimensions predicted by string theory in 
order to successfully explain the origins of a large, $3 + 1$ dimensional universe.  
Brandenberger and Vafa
proposed that, from a cosmological perspective, it is much more logical to consider
the \it decompactification \rm of three of the spatial directions.  In other words,
one starts in a universe with nine dimensions, each compactified close to the Planck length
and then, for one reason or another, three spatial dimensions grow large.  

The toy model of the early
universe considered here is a nine dimensional box with each dimension having equal length,
$R$.  The box is filled with strings and periodic boundary conditions
are imposed as described
in Section (\ref{duality}).  

In the SBB model
it is possible to plot the scale factor $R$ vs. $t$ using 
the Einstein equations (\fig{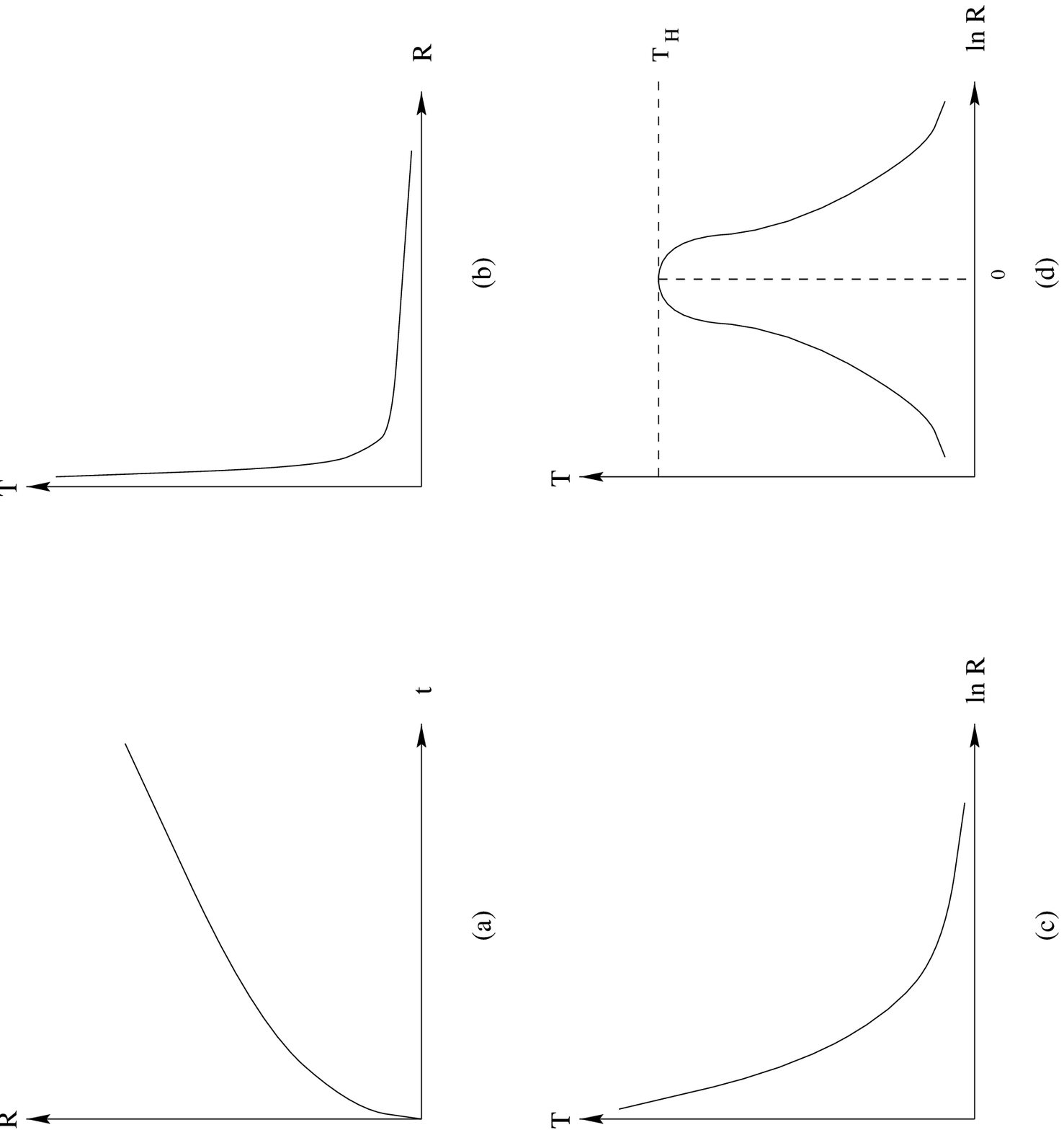}(a)).  
For the radiation dominated epoch,
$R \propto t^{1/2}$.  Furthermore, it is possible to plot $R$ vs. the temperature $T$, where $T \propto 1/R$ (\fig{rhb.eps} (b) and (c)).
In string theory we have no analogue of Einstein's equations and hence we cannot
obtain a plot of the scale factor, $R$ vs. $t$.  On the other hand, we do know the entire spectrum 
of string states and so we \it{can} \rm obtain an analogue of the 
$R$ vs. $T$ curve (see \fig{rhb.eps}(d)).  Note that the region of \fig{rhb.eps}(d) near
the Hagedorn temperature is not well understood, and canonical ensemble approaches 
break down.  Fortunately, the regions to the left and right of $T_H$ are connected via dualities.
The interested reader should see e.g. \cite{ref:hagedorn1965}-\cite{ref:meanaet1999} for 
more modern investigations of the Hagedorn transition.

Recall, that in General Relativity the temperature $T$ goes 
to infinity as the radius $R$ decreases.
As we have already mentioned, string theory predicts a maximum temperature,
$T_H$ and therefore one should expect the stringy $R$ vs. $T$ curve to be drastically altered.
Furthermore, we found that string theory enjoys the $R \rightarrow 1/R$ symmetry 
which leads
to a $\ln R \rightarrow - \ln R$ symmetry in \fig{rhb.eps}(d).  
For large values of $R$,
$R \propto 1/T$ is valid since the winding modes are irrelevant 
and the theory looks like a point particle theory.  For small $R$
the $T-R$ curve begins to flatten out, approach the Hagedorn temperature and then 
as we continue to go to smaller values of $R$ the temperature begins to \it decrease\rm.
This behavior is a consequence of the T-duality of string theory.  As $R$ shrinks,
the winding modes which are absent in point particle theories become lighter and lighter,
and are therefore easier to produce.  Eventually, (with entropy constant) the thermal
bath will consist mostly of winding modes, which explains the decrease in temperature
once one continues past $T_H$ to smaller values of $R$.  
\newpage
\vspace{.5cm}
\hglue 2cm
\psfig{figure=rhb.eps,height=11cm,angle=-90}\label{rhb.eps}
\begin{quote}
\scriptsize Figure \ref{rhb.eps}: In (a) (and (b)), we have plotted $R$ vs. $t$
($T$ vs. $R$) for the SBB model.  Figures (c) and (d) are plots
of $T$ vs. $\ln R$ for both the SBB and String cosmological models
respectively.  Note the $\ln R \rightarrow - \ln R$ symmetry in (d). \end{quote}

An observer traveling from large $R$ to small $R$,  actually
sees the radius contracting to $R=1$ (in Planck units) and then \it expanding \rm
again.  This makes us more comfortable with the idea of the temperature beginning
to decrease after $R=1$.  The reason for this behavior is that 
the observer must modify
the measuring apparatus to measure distance in terms of light states. The details
for making this change of variables are described in \cite{ref:brandenberger1989}.

Hence, the observer described above 
encounters an oscillation of the universe.  This encourages one
to search for cosmological solutions in string theory
where the universe oscillates from small to large, eliminating the
initial and final singularities found in (SBB) models.

\subsection{The Dimensionality Problem} \label{dimension}

We are now ready to ask the question, how can superstring theory, a theory consistently
formulated in ten dimensions give rise to a universe with only four macroscopic dimensions?
This is equivalent within the context of our toy model to asking why should three of the nine spatial dimensions
of our box ``want" to expand? To address this question, note the following
observation: winding modes lead to negative pressure in the thermal bath.  
To understand this, recall that as the volume of the box increases, the energy in the
winding modes also increases (\ref{eq:winding}).  Thus the phase space available to the
winding modes
decreases, which brings us to the conclusion that winding modes would ``like" to 
prevent expansion.  The point is that it costs a lot of energy to expand with
winding modes around.  Thermal equilibrium demands that the number of
winding modes must decrease as $R$ increases (since the winding modes become heavier).
Therefore, we conclude that expansion can only occur when the system is in
thermal equilibrium, which favors fewer of the winding states as $R$ increases.
If, on the other hand, the winding modes are not in thermal equilibrium they will
become plentiful and thus any expansion will be slowed and eventually brought
to a halt.

Thermal equilibrium of the winding modes requires string interactions of the form
\be\label{eq:equil}
W + \bar W \Leftrightarrow unwound \,\,\, states
\,.
\ee
Here $W$ is a winding state and $\bar W$ is a winding state with opposite winding
as depicted in \fig{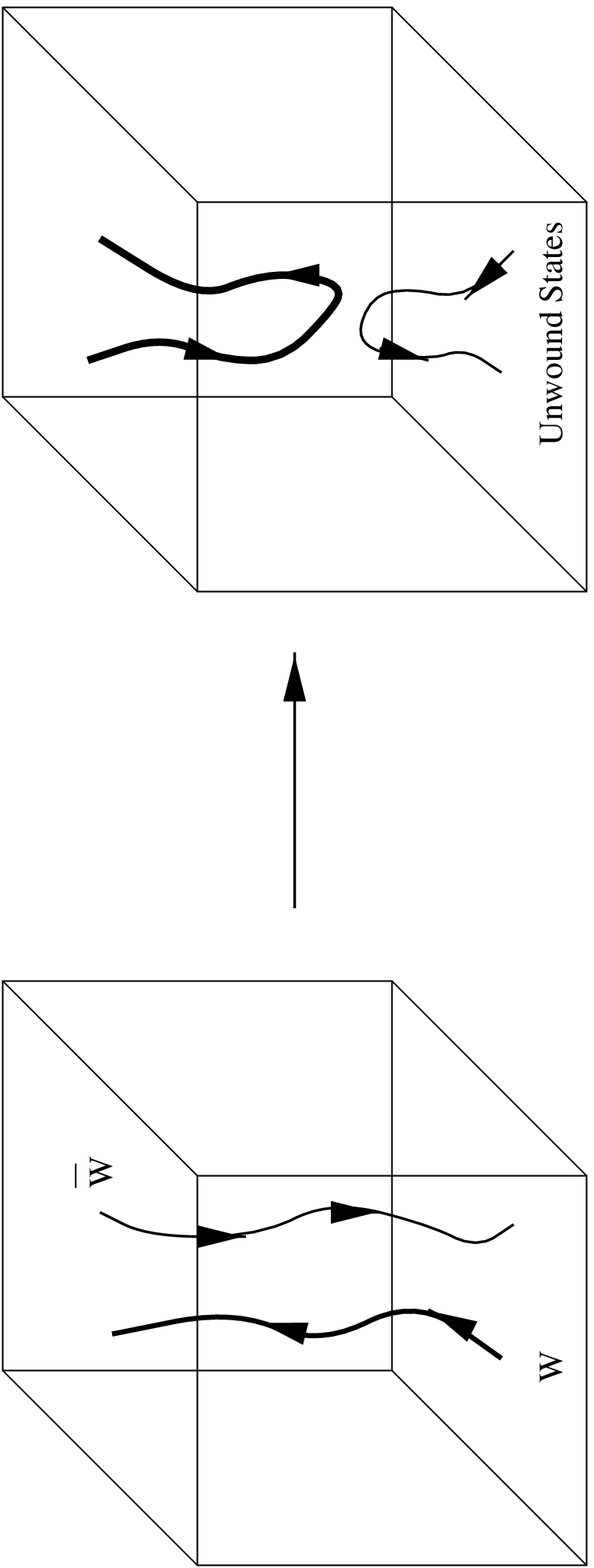}.  

\vspace{.75cm}
\hglue -.5cm
\psfig{figure=wind.eps,height=5.5cm,width=15cm,angle=-90}\label{wind.eps}
\begin{quote}
\scriptsize Figure \ref{wind.eps}: Strings that interact with opposite
windings become unwound states.
\end{quote}

In order for such processes to occur, the strings must
come to within a Planck length of one another.  As the winding strings
move through spacetime they span out two dimensional world sheets.  In order to 
interact, their worldsheets must intersect, but in a nine dimensional box
the strings will probably not intersect because $2+2 < 9+1$.  Since there is
so much room in the box, the strings will have a hard time finding
one another in order for their worldsheets to intersect and therefore it is
unlikely that they will unwind.
If the winding strings do not unwind, and the box starts to expand, the
winding states will fall out of thermal equilibrium and the expansion will be halted.

The conclusion is that the largest spacetime dimensionality consistent
with maintaining thermal equilibrium is four. Since, $2+2 = 3+1$, and therefore
the largest number of spatial dimensions which can expand is three.
In the next section we will see how this scenario can be incorporated
into our current understanding of string theory.

\subsection{Brane Gases and the ABE Mechanism}

Recent developments in M/string theory have revealed that strings are not
the only fundamental degrees of freedom in the theory.  The spectrum
of fundamental states also includes higher dimensional extended
objects known as D-branes.  Here we will examine the way in which the
BV scenario unfolds in the presence of D-branes in the early 
universe as constructed by Alexander, Brandenberger and Easson (ABE) \cite{ref:alexanderet}.  
Specifically, we are interested
in finding out if the inclusion of branes affects the 
cosmological implications of \cite{ref:brandenberger1989}.  Note that 
this approach to string cosmology is in close analogy with the starting point
of the standard big-bang model and is very different from other cosmological models
which have attempted to include D-branes, for example the brane-world scenarios 
discussed in Sections \ref{heterotic} and \ref{led}.  However, possible relations between this
model and brane-world scenarios will be discussed later.

Our initial state will be similar to that of
\cite{ref:brandenberger1989}.  We assume that the universe started
out close to the Planck length, dense and hot and with all degrees of
freedom in thermal equilibrium.  As in \cite{ref:brandenberger1989}, we choose a toroidal 
geometry in all spatial dimensions.  The initial state will
be a gas composed of the fundamental branes in the theory.  We will
consider 11-dimensional M-theory compactified on $S^1$ to yield
10-dimensional Type II-A string theory.  The low-energy effective
theory is supersymmetrized dilaton gravity.  Since M-theory admits
the graviton, 2-branes and 5-branes as
fundamental degrees of freedom, upon the $S^1$ compactification
we obtain 0-branes, strings (1-branes), 2-branes, 4-branes,
5-branes, 6-branes and 8-branes in the 10-dimensional universe.

The details of the compactification will not be discussed here,
however we will briefly mention the origins of the above objects
from the fundamental eleven-dimensional, M-theory perspective.
The 0-branes of the II-A theory are the BPS states of 
nonvanishing $p_{10}$.  In M-theory these are the states of
the massless graviton multiplet.  The 1-brane of the II-A theory
is the fundamental II-A string which is obtained by wrapping
the M-theory supermembrane around the $S_1$.  The 2-brane
is just the transverse M2-brane.  The 4-branes are wrapped
M5-branes.  The 5-brane of the II-A theory is a solution 
carrying magnetic NS-NS charge and is an M5-brane that is
transverse to the eleventh dimension.  The 6-brane field
strength is dual to that of the 0-brane, and is a KK
magnetic monopole.  The 8-brane is a source for
the dilaton field \cite{ref:polchinski1998}.  

The low-energy bulk effective action for the above setup is
\begin{eqnarray} \label{bulk}
S_{bulk} \, = \, {1 \over {2 \kappa^2}} \int d^{10}x \sqrt{-G} e^{-2 \phi} \bigl[ R &+& 4 G^{\mu \nu} \nabla_\mu \phi \nabla_\nu \phi \nonumber \\
&-& {1 \over {12}} H_{\mu \nu \alpha}H^{\mu \nu \alpha} \bigr] \, ,
\end{eqnarray}
where $G$ is the determinant of the background metric $G_{\mu \nu}$,
$\phi$ is the dilaton, $H$ denotes the field strength corresponding to
the bulk antisymmetric tensor field $B_{\mu \nu}$, and $\kappa$ is
determined by the 10-dimensional Newton constant in the usual way.

For an individual $p$-brane
the action is of the Dirac-Born-Infeld form
\begin{equation} \label{brane}
S_p \, = \, T_p \int d^{p + 1} \zeta e^{- \phi} \sqrt{- det(g_{mn} + b_{mn} + 2 \pi \alpha' F_{mn})}
\end{equation}
where $T_p$ is the tension of the brane, $g_{mn}$ is the induced metric on the brane, $b_{mn}$ is the 
induced antisymmetric tensor field, and $F_{mn}$ the field strength tensor of gauge fields $A_m$ living on the brane. 
The total action is the sum of the 
bulk action (\ref{bulk}) and the sum of all of the brane actions
(\ref{brane}), each coupled as a delta function source (a delta
function in the directions transverse to the brane) to the
10-dimensional action.

In the string frame the tension of a $p$-brane is
\begin{equation}
T_p \, = \, {{\pi} \over {g_s}} (4 \pi^2 \alpha')^{-(p + 1)/2} \, ,
\end{equation}
where $\alpha' \sim l_{st}^2$ is given by the string length scale
$l_{st}$ and $g_s$ is the string coupling constant.

In order to discuss the dynamics of this system, we
will need to compute the equation of state for the brane gases for
various $p$.  There are three types of modes that we will need to 
consider.  First, there are the winding modes.  The background space
is $T^9$, and hence a $p$-brane can wrap around any set
of $p$ toroidal directions.  These modes are related by t-duality to
the momentum modes corresponding to center of mass motion of the
branes.  Finally, the modes corresponding to fluctuations of the branes
in the transverse directions are (in the low-energy limit) described
by scalar fields on the brane, $\phi_i$.  There are also bulk matter
fields and brane matter fields.

We are mainly interested in the effects of winding modes and
transverse fluctuations to the evolution of the universe, and
therefore we will neglect the antisymmetric tensor field $B_{\mu\nu}$.
We will take our background metric with conformal time $\eta$ to be
\begin{equation}
G_{\mu \nu} \, = \, a(\eta)^2 diag(-1, 1, ..., 1) \, ,
\end{equation}
where $a(\eta)$ is the cosmological scale factor.

If the transverse fluctuations of the brane and the gauge fields
on the brane are small, the brane action can be expanded as
\begin{eqnarray} \label{actexp}
S_{brane} \, &=& \, T_p \int d^{p+1} \zeta a(\eta)^{p + 1} e^{-\phi} \nonumber \\
& & e^{{1 \over 2} tr log(1 + \partial_m \phi_i \partial_n \phi_i + a(\eta)^{-2} 2 \pi \alpha' F_{mn})} \nonumber \\
&=& \, T_p \int d^{p + 1} \zeta a(\eta)^{p + 1} e^{- \phi} \\
& & (1 + {1 \over 2} (\partial_m \phi_i)^2 - \pi^2 {\alpha'}^2 a^{-4} F_{mn}F^{mn}) \, . \nonumber
\end{eqnarray}
The first term in the parentheses in the last line represents the
brane winding modes, the second term corresponds to the transverse fluctuations, and the third term to brane matter.  
In the low-energy limit, the transverse fluctuations of the 
brane are described by a free scalar field action, and the
longitudinal fluctuations are given by a Yang-Mills theory. The
induced equation of state has pressure $p \geq 0$ \footnote{Note that
the above result is still valid when brane fluctuations and fields are large \cite{ref:alexanderet} .}.

To find the equation of state for the winding modes, we use equation
(\ref{actexp}) to get
\begin{equation} \label{EOSwind}
\tilde{p} \, = \, w_p \rho \,\,\, {\rm with} \,\,\, w_p = - {p \over d}
\end{equation}
where $d$ is the number of spatial dimensions (9 in our case), and where $\tilde{p}$ and $\rho$ stand for the 
pressure and energy density, respectively.

Fluctuations of the branes and brane matter are given by free scalar
and gauge fields on the branes.  These may be viewed as particles in
the transverse directions extended in brane directions.  Therefore,
the equation of state is simply that of ordinary matter,
\begin{equation} \label{EOSnw}
\tilde{p} \, = \, w \rho \,\,\, {\rm with} \,\,\, 0 \leq w \leq 1 \, .
\end{equation} 
From the action (\ref{actexp}) we see that the energy in the winding
modes will be
\begin{equation} \label{winden}
E_p(a) \, \sim \, T_p a(\eta)^p \, ,
\end{equation}
where the constant of proportionality is dependent on the number of
branes.

The equations of motion for the background are given by \cite{ref:tseytlinet1992,ref:veneziano1991}
\begin{eqnarray} \label{EOMback1}
- d \dot{\lambda}^2 + \dot{\varphi}^2 \, &=& \, e^{\varphi} E \\
\label{EOMback2}
\ddot{\lambda} - \dot{\varphi} \dot{\lambda} \, &=& \, {1 \over 2} e^{\varphi} P \\
\label{EOMback3}
\ddot{\varphi} - d \dot{\lambda}^2 \, &=& \, {1 \over 2} e^{\varphi} E \, ,
\end{eqnarray}
where $E$ and $P$ denote the total energy and pressure, respectively,
\begin{equation}
\lambda(t) \, = \, log (a(t)) \, ,
\end{equation}
and $\varphi$ is a shifted dilaton field which absorbs the space volume factor
\begin{equation}
\varphi \, = \, 2 \phi - d \lambda \, .
\end{equation}
The matter sources $E$ and $P$ are made up of all the components of the brane gas:
\begin{eqnarray}
E \, &=& \, \sum_p E_p^w + E^{nw} \nonumber \\
P \, &=& \, \sum_p w_p E_p^w + w E^{nw} \, ,
\end{eqnarray}
where the superscripts $w$ and $nw$ stand for the winding modes and
the non-winding modes, respectively.  The contributions of the
non-winding modes of all branes have been combined into one term. The
constants $w_p$ and $w$ are given by (\ref{EOSwind}) and
(\ref{EOSnw}). Each $E_p^w$ is the sum of the energies of all of the
brane windings with fixed $p$.

We may now draw the comparison between the ABE mechanism and
\cite{ref:brandenberger1989}.  First of all we see that
both t-duality and limiting Hagedorn temperature are still manifest
once we include the $p$-branes \cite{ref:alexanderet}.  Therefore,
there is no physical singularity as $R \rightarrow 0$.  What about the
de-compactification mechanism described in section (\ref{early})?
Recall that our initial conditions are in a hot, dense regime
near the self dual point $R=1$.  All the modes (winding, oscillatory and momentum) of all
the $p$-branes will be excited.  By symmetry, we assume that there are
equal numbers of winding and anti-winding modes in the system and hence
the total winding numbers cancel as in \cite{ref:brandenberger1989}.  

Now assume that the universe begins to expand in all directions.  
The total energy in the winding modes increases with $\lambda$ as
(\ref{winden}), so the largest $p$-branes contribute the most.
The classical counting argument discussed in \cite{ref:brandenberger1989}
is easily generalized to our model.  When winding modes meet
anti-winding modes, the branes unwind (recall \fig{wind.eps}) and allow a certain number
of dimensions to grow large.

Consider the probability that the world-volumes of two $p$-branes
in spacetime will intersect.  The winding modes of $p$-branes are likely to
interact in at most $2p+1$ spatial dimensions. \footnote{To see this, consider the example of two
particles (0-branes) moving through a space of dimension $d$.  These particles
will definitely interact (assuming the space is periodic) if $d=1$, whereas they
probably will not find each other in a space with $d>1$.}

Since we are in
$d=9$ spatial dimensions the $p = 8,\, 6,\, 5,\, 4$ branes will interact
and hence unwind very quickly.  For $p < 4$ a hierarchy of dimensions will
be allowed to grow large.  Since the energy contained in the
winding modes of $2$ branes is larger than that of strings 
(see (\ref{winden})) the $2$ branes will have an important effect first.
The membranes will allow a $T^5$ subspace to grow large.  Within
this $5$ dimensional space the $1$-branes will allow a
$T^3$ subspace to become large.  We therefore reach the conclusion
that the inclusion of D-branes into the spectrum of fundamental
objects in the theory will cause a hierarchy of subspaces to become large
while maintaining the results of the BV scenario, explaining the origin of our
$3+1$-dimensional universe. 

Let us summarize the evolution of the ABE universe.  The universe starts out in an initial state
close to the self-dual point ($R=1$), a 9-dimensional toroidal
space, hot, dense and
filled with particles, strings and $p$-brane gases.  The universe
then starts to expand according to the background equations
of motion (\ref{EOMback1} - \ref{EOMback3}).  Branes with the
largest value of $p$ will have an effect first,
and space can only expand further if the winding modes annihilate.
The $8,\,6,\,5$ and $4$-branes winding modes annihilate quickly, followed
by the $2$-branes which allow only $5$ spatial dimensions to 
become large.  In this $T^5$ the strings allow a $3$-dimensional
subspace to become large.  Hence, it is reasonable to hypothesize the
existence of a $5+1$-dimensional effective theory at some point
in the early history of the universe.  In particular, one is 
tempted to draw a relation between this $5$-dimensional picture
and the scenario of large extra dimensions proposed in \cite{ref:dvaliet1999a}.

There are several problems with the toy model analyzed above.  Most of these
have already been mentioned by the authors.   First, the strings and
branes are treated
classically.  Quantum effects will cause the strings to take
on a small but finite thickness \cite{ref:karlineret1988}, although in
our case we are restricted to energy densities lower than the typical string
density, and hence the effective width of the strings is of string scale \cite{ref:sen1999}.    
This presumably will also apply to the branes, although there
is no current, consistent quantization scheme developed for branes.

In this scenario there is a \it brane problem \rm.  This is a new problem for
cosmological theories with stable branes analogous to the domain wall problem
in cosmological scenarios based on quantum field theories with stable
domain walls.  However, we have found background solutions in our models which
approach a point of loitering \cite{ref:sahniet1992}.  Loitering occurs if at some point in the evolution
of the universe the size of the Hubble radius
extends larger than the physical radius. Such a phase in the background cosmological
evolution will naturally solve the brane problem.    
  
The toroidal topology of the 
compactified manifold was chosen for simplicity.  It is important from 
the point of view
of string theory to consider how things would change if this manifold
was a Calabi-Yau space.  Calabi-Yau three-folds do not admit one cycles for strings to wrap around, 
although they are necessary if the four-dimensional low energy effective theory is to have $N=1$
supersymmetry.  Note that in cosmology we do not necessarily expect $N=1$ supersymmetry.  
In particular, maximal supersymmetry is consistent with the toriodal background used.

Also, it was argued in \cite{ref:hull1999} that
M-theory should not be formulated in a spacetime of definite dimension
or {\it signature}.  In other words, we must ultimately be able to explain
why there is only one time dimension.

Although there is no horizon problem present in this scenario since
the universe starts out near the string length and hence there
are no causally disconnected regions of space, other problems
solved by inflation such as the flatness and structure formation
problems are still present.  Other less significant concerns are stressed in \cite{ref:alexanderet}.
This scenario provides a new method for studying string cosmology which is similar to the 
SBB model and utilizes $p$-branes in a very different way from scenarios 
involving large extra dimensions.

\section{Pre-Big-Bang}\label{PBB}

The next attempt to marry cosmology with string theory we will review
was proposed in the early 1990s by Veneziano and Gasperini 
\cite{ref:veneziano1991}-\cite{ref:veneziano2000}.  

\subsection{Introduction}\label{intropbb}

The Pre-Big-Bang (PBB) model \footnote{For an updated collection of papers on this
model see http://www.to.infn.it/$\sim$gasperin.} 
is based on the low energy effective action of string theory,
which in $d$ spatial dimensions is given by
\be\label{eq:lea}
S = - \frac{1}{2\l_s^{d-1}} \int d^{d+1}x \, \sqrt{-g} \, e^{-\varphi} \, 
\left[ R + (\d_\m \varphi)^2 + \cdots \, \right]
\,,
\ee
where $\varphi$ is the dilaton and $\l_s$ is the string length scale.  The qualitative
differences between the PBB model, and the SBB model based on the Einstein-Hilbert
action,
\be\label{eq:eh}
S = - \frac{1}{2\l_p^{d-1}} \int d^{d+1}x \, \sqrt{-g} \, R
\,,
\ee
are most easily visualized by plotting the history of the curvature of the universe
(see \fig{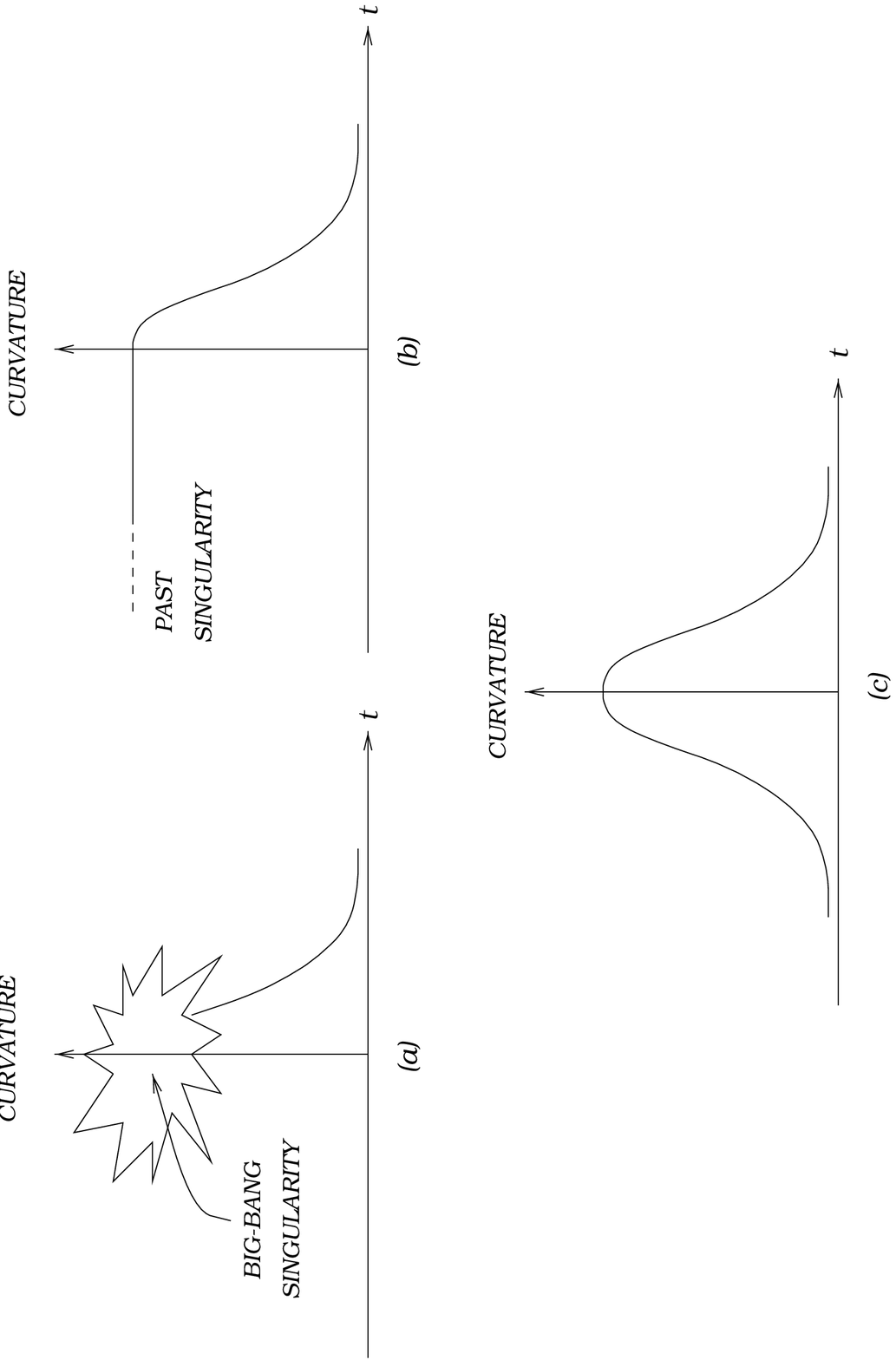}) according to each theory.  In the SBB scenario the curvature increases as
we go back in time, eventually reaching an infinite value at the Big-Bang singularity.
In standard inflationary models the curvature reaches some fixed value as $t$ decreases
at which point the universe enters a de Sitter phase.  It has been shown however that
such an inflationary phase cannot last forever, for reasons of geodesic completeness,
and that the initial singularity problem still remains 
\cite{ref:bordeet1993,ref:brandenberger1999}.  
The cosmology generated by (\ref{eq:lea}) differs
drastically from the standard scenarios.  The action (\ref{eq:lea}) without
the ``$\cdots$'' terms does not realize the PBB scenario, as we will
discuss below.  In the PBB model, as one travels back in time the curvature increases
as in the previously mentioned models, but in the PBB a maximum curvature is reached at which point
the curvature and temperature actually begin to {\it decrease}.  
Although we will examine the details
of how this occurs below, a few simple considerations make us feel more comfortable
with this picture.

For one, string theory predicts a natural cut-off length scale,
\be\label{eq:lstring}
\l_s = \sqrt{\frac{\hbar}{T}} \sim 10 \, l_{pl} \sim 10^{-32} \cm
\,,
\ee
where $T$ is the string tension and $l_{pl}$ is the Planck length.  So it is
natural from the point of view of strings to expect a maximum possible curvature.
Logically, as we travel back in time there are only two possibilities if we want
to avoid the initial singularity.  Either the curvature starts to grow again
before the de Sitter phase,
in which case we are still left with a singularity shifted earlier in time,
or the curvature begins to decrease again, which is what happens in the PBB scenario (\fig{pbb.eps}c).  
This behavior is a consequence of scale-factor duality.

\vspace{.75cm}
\hglue 1cm
\psfig{figure=pbb.eps,height=8.5cm,angle=-90}\label{pbb.eps}
\begin{quote}
\scriptsize Figure \ref{pbb.eps}: Curvature plotted versus time for,
(a) the SBB model, (b) the standard inflationary model and (c) the PBB scenario.
\end{quote}

\subsection{More on Duality}\label{mored}

To demonstrate the enhanced symmetries present in the PBB model we will examine the
consequences of scale-factor duality.  The Einstein-Hilbert action (\ref{eq:eh}) is invariant under
time reversal.  Hence, for every solution $a(t)$ there exists a solution $a(-t)$.  Or
in terms of the Hubble parameter $H(t) = \dot a(t) / a(t)$, for every solution
$H(t)$ there exists a solution $- H(-t)$.  Thus, if there is a solution representing
a universe with decelerated expansion and decreasing curvature ($H>0$, \, $\dot H < 0$) there
is a ``mirror" solution corresponding to a contracting universe ($H(-t)$, \, $H < 0$).

The action of string theory (\ref{eq:lea}) is not only invariant under time reversal,
but also under inversion of the scale factor $a(t)$, (with an appropriate transformation
of the dilaton).  For every cosmological solution $a(t)$ there is a solution $\tilde a = 1/a(t)$,
provided the dilaton is rescaled, $\varphi \rightarrow \tilde \varphi = \varphi - 2d\, \ln a$.  
Hence, time reversal symmetry
together with scale-factor duality imply that every cosmological solution has four branches, 
\fig{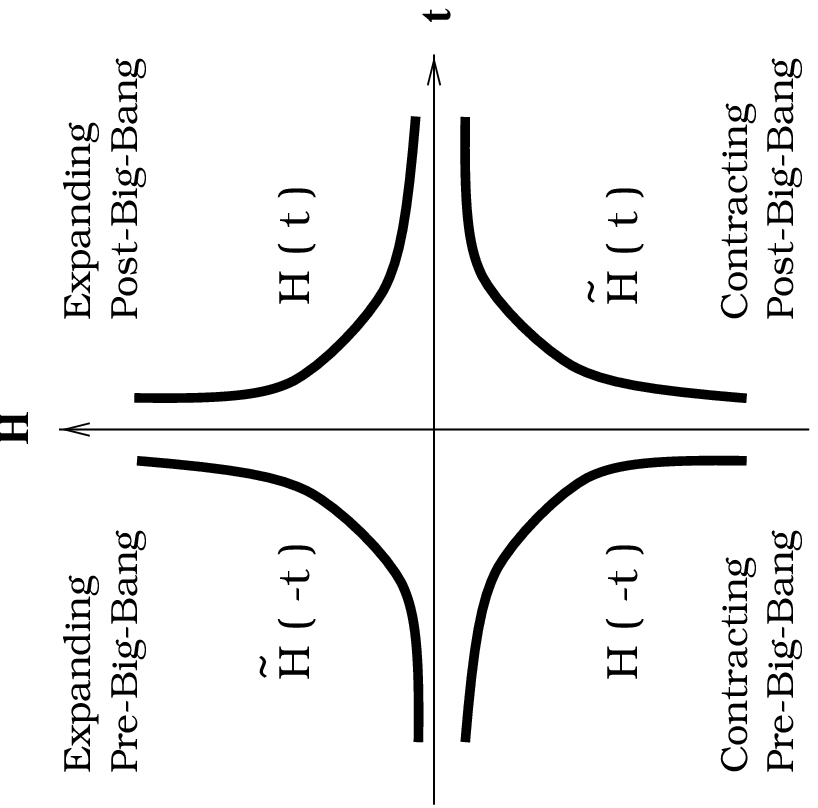}.
For the standard scenario of decelerated expansion and decreasing curvature
($H(t) > 0$, \, $\dot H(t) < 0$ ) there is a dual partner
solution describing a universe with accelerated expansion parameter $\tilde H(t)$ and growing
curvature $\dot {\tilde H}(- t)$.

\vspace{.75cm}
\hglue 3cm
\psfig{figure=branch.eps,height=8cm,angle=-90}\label{branch.eps}
\begin{quote}
\scriptsize Figure \ref{branch.eps}: The four branches of a string cosmological
solution resulting from scale-factor duality and time reversal.
\end{quote}

We will now show how one can create a universe from the string theory perturbative
vacuum, that today looks like the 
standard cosmology.  This problem is analogous to
finding a smooth way to connect the Pre-Big-Bang phase with a Post-Big-Bang phase, or how
to successfully connect the upper-left side of \fig{branch.eps} to the upper-right 
side.  In general,
the two branches are separated by a future/past singularity and it appears that
in order to smoothly connect the branches of growing and decreasing curvature one requires
the presence of higher order loop and/or derivative corrections to the effective action
(\ref{eq:lea}).  This cancer of the PBB model is know as the 
Graceful Exit Problem (GEP) and is the subject of many research papers 
(see \cite{ref:veneziano2000,ref:lidseyet1999} for a collection of
references).  

One example of how the GEP can be solved is given in \cite{ref:eassonet1999}.
In this work we consider a theory obtained by adding to the usual
string frame dilaton gravity action specially constructed higher 
derivative terms motivated by the limited curvature construction of
\cite{ref:mukhanovet1992}.  The action is (\ref{eq:lea}) with
the ``$\cdots$'' term being replaced by the constructed higher derivative terms.  
In this scenario all solutions of
the resulting theory of gravity are nonsingular and for initial
conditions inspired by the PBB scenario solutions exist which
smoothly connect a ``superinflationary'' phase with $\dot H > 0$, to an expanding
FRW phase with $\dot H<0$, solving the GEP in a natural way.

\subsection{PBB-Cosmology}\label{pbbcos}

Here we examine cosmological solutions of the PBB model.  By adding
matter in the form of a perfect fluid to the effective action (\ref{eq:lea}) 
(without the ``$\cdots$'' terms) and taking
a Friedmann-Robertson-Walker background with $d=3$, we vary the action to get
the equations of motion for string cosmology,
\begin{eqnarray}\label{eq:eomstring}
        \dot\vp^2 - 6H\dot\vp + 6H^2 & = & e^\vp \r \,, \\
        \dot H - H\dot\vp + 3H^2 & = & \frac{1}{2} e^\vp p \,,\nonumber \\
        2 \ddot \vp + 6H\dot\vp - \dot\vp^2 - 6\dot H - 12H^2 & = & 0 \nonumber 
\,.
\end{eqnarray}
As an example, for $p=\r/3$ the equations with constant dilaton are exactly solved
by
\be\label{eq:rad}
a \propto t^{1/2}, \qquad \r \propto a^{-4}, \qquad \vp = \mbox{const.}
\,,
\ee
which is the standard scenario for the radiation dominated epoch, having
decreasing curvature and decelerated expansion:
\be\label{eq:raddom}
\dot a > 0, \qquad \ddot a < 0, \qquad \dot H < 0
\,.
\ee
But there is also a solution obtained from the above via time translation and 
scale-factor duality,
\be\label{eq:trans}
t \rightarrow -t, \qquad a \propto (-t)^{-1/2}, \qquad \vp \propto -3 \ln(-t),
\qquad \r = -3p \propto a^{-2} \,.
\ee
This solution corresponds to an accelerated, inflationary expansion, with
growing dilaton and growing curvature:
\be\label{eq:supinf}
\dot a > 0, \qquad \ddot a >0, \qquad \dot H > 0 \,.
\ee
Solutions with such behavior are called ``superinflationary" and are
located in the upper left quadrant of \fig{branch.eps}.

Let us briefly review the history of the universe 
as predicted by the PBB scenario.  Recall, that in the SBB model the universe
starts out in a hot, dense and highly curved regime.  In contrast, the PBB
universe has its origins in the simplest possible state we can think of, namely
the string perturbative vacuum.  Here the universe consists only of a sea of
dilaton and gravitational waves.  It is empty, cold and flat, which means that
we can still trust calculations done with the classical, low-energy effective
action of string theory.  

In \cite{ref:kaloperet1999}, the authors showed
that in a generic case of the PBB scenario, the universe at the
onset of inflation must already be extremely large and homogeneous.  In order
for inflation to solve flatness problems the initial size
of a homogeneous part of the universe before PBB inflation
must be greater than $10^{19}l_s$.  In response, it was proposed in 
\cite{ref:buonannoet1998} that the initial state of the PBB model
is a generic perturbative solution of the tree-level, low-energy effective
action.  Presumably, quantum fluctuations 
lead to the formation of many black holes (\fig{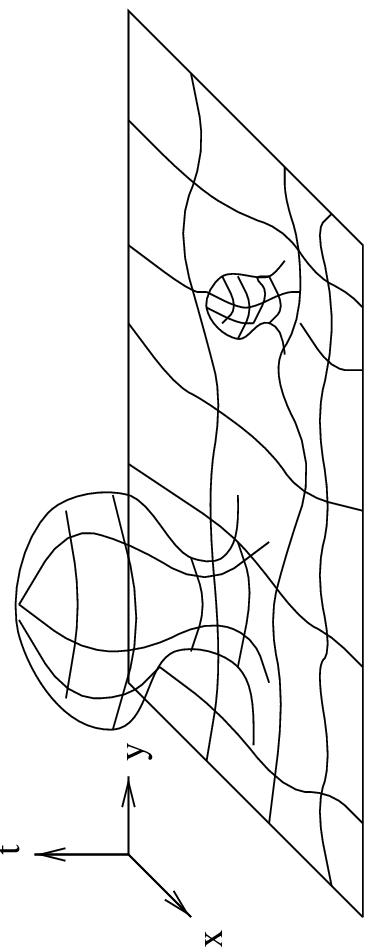}) in the gravi-dilaton 
sector (in the Einstein frame).  Each such singular space-like hypersurface of 
gravitational collapse becomes
a superinflationary phase in the string frame 
\cite{ref:feinsteinet2000, ref:ghoshet2000,ref:bozzaet2000,ref:buonannoet1998}.  
After the period of dilaton-driven inflation the universe evolves in accordance 
with the SBB model.

\vspace{.5cm}
\hglue 2cm
\psfig{figure=bubbl.eps,height=4cm,angle=-90}\label{bubbl.eps}
\begin{quote}
\scriptsize Figure \ref{bubbl.eps}: A $2+1$ dimensional slice of the string
perturbative vacuum giving rise to black hole formation in the Einstein frame.
\end{quote}

To conclude let us mention a few benefits of
the PBB scenario.  For one, there is no need to {\it invent} inflation, or fine 
tune a potential for the inflaton.  This model provides a ``stringy" realization of
inflation which sets in naturally and is dilaton 
driven.  Pair creation (quantum instabilities) 
provides a mechanism to heat up an initially cold
universe in order to produce a hot big-bang with homogeneity, isotropy and flatness.
This scenario also has observable consequences.

Problems with this scenario include the graceful exit problem, mentioned above.
This is the problem of smoothly connecting the phases of growing and decreasing
curvature, a process that is not well understood and requires further investigation.
Most cosmological models require a potential for the dilaton to 
be introduced by hand in order to
freeze the dilaton at late times.  In general it is believed that the dilaton should
be massive today, otherwise we would notice its effects on physical gauge couplings.

Inclusion of a non-vanishing $B_{\mu\nu}$ into the action (\ref{eq:lea}) greatly reduces
the initial conditions which give rise to inflation \cite{ref:lidseyet1999}.  
Also the initial collapsing region must be sufficiently large and weakly coupled.
Lastly, the dimensionality problem is still present in this model.

\section{Cosmology and Heterotic M-Theory}\label{heterotic}

In this section we will focus on the work of Lukas, Ovrut and Waldram 
(LOW)\cite{ref:lukaset1998}
in 1998,
which is
based on the heterotic M-theory of Ho\v{r}ava and Witten \cite{ref:horavaet1996a,
ref:horavaet1996b,ref:witten1996}.  Their motivation was to see if it is possible
to construct a realistic, cosmological model starting from the most fundamental
theory we know.

\subsection{Ho\v{r}ava-Witten Theory}

In 1996, Ho\v{r}ava and Witten showed that eleven-dimensional M-theory compactified
on an $S^1/Z_2$ orbifold with a set of $E_8$ gauge supermultiplets
on each ten-dimensional orbifold fixed plane can be identified with 
strongly coupled $E_8 \times E_8$ heterotic string theory\cite{ref:horavaet1996a,
ref:horavaet1996b}.  The basic setup is that of \fig{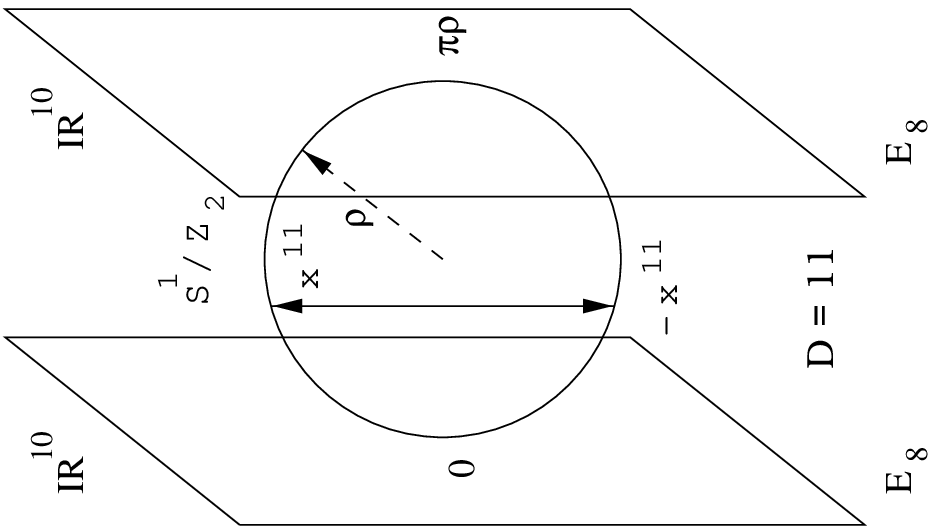}, where the orbifold
is in the $x^{11}$ direction and $x^{11} \in [-\pi \r,\pi \r]$ with the endpoints
being identified.  The orbifolding with $Z_2$ leads to the symmetry 
$x^{11} \rightarrow - x^{11}$. It has been shown that 
this M-theory limit can be consistently compactified on a deformed Calabi-Yau
three-fold resulting in an $N=1$ supersymmetric theory in four dimensions 
(see fig.(5.2)).
In order to match (at tree level) the gravitational and grand-unified gauge couplings
one finds the requirement $R_{orb} > R_{CY}$, where $R_{orb}$ is the radius of the orbifold
and $R_{CY} \approx 10^{16} \,\GeV$ is the radius of the Calabi-Yau space.
This picture leads to the conclusion that the universe may have gone
through a phase in which it was effectively five-dimensional, and therefore
provides us with a previously unexplored regime in which to study the early universe.

\vspace{.5cm}
\hglue 5cm
\psfig{figure=daec.eps,height=7.5cm,angle=-90}\label{daec.eps}
\begin{quote}
\scriptsize Figure \ref{daec.eps}:  The Ho\v{r}ava-Witten scenario.  
One of the eleven-dimensions has been compactified onto the orbifold $S^1/Z_2$.  The 
manifold is ${\mathcal M} = I\!\!R^{10} \times S^1/Z_2$.
\end{quote}

Here we construct the five-dimensional effective 
theory via reduction
of Ho\v{r}ava-Witten on a Calabi-Yau three-fold, and then
show how this can lead to a four-dimensional toy model for a Friedmann-Robertson-Walker (FRW)
universe.

We start with an eleven-dimensional action with bosonic contribution
\be\label{eq:action}
S = S_{SUGRA} + S_{YM}
\,,
\ee
where $S_{SUGRA}$ is the action of eleven-dimensional supergravity
\begin{eqnarray}\label{eq:asugra}
S_{SUGRA} & = & -\frac{1}{2\kappa^2} \int_{M^{11}} \sqrt{- g} \, \bigg[ R + 
\frac{1}{24} G_{IJKL}G^{IJKL} \nonumber \\
        & & + \frac{\sqrt{2}}{1728} \epsilon^{I_1 \cdots I_{11}} C_{I_1 I_2 I_3} G_{I_4 \cdots I_7}
        G_{I_8 \cdots I_{11}} \bigg]
\,,
\end{eqnarray}  
and $S_{YM}$ are two $E_8$ Yang-Mills theories on the ten-dimensional orbifold planes
\begin{eqnarray}\label{eq:aym}
S_{YM} & = & -\frac{1}{8 \pi \kappa^2} \left( \frac{\kappa}{4\pi} \right)^{2/3}
        \int_{M^{(1)}_{10}} \sqrt{- g} \, \left\{ tr(F^{(1)})^2 - \frac{1}{2} tr R^2 \right\} 
        \nonumber \\
        & & -\frac{1}{8 \pi \kappa^2} \left( \frac{\kappa}{4\pi} \right)^{2/3}
        \int_{M^{(2)}_{10}} \sqrt{- g} \, \left\{ tr(F^{(2)})^2 - \frac{1}{2} tr R^2 \right\}
\,.
\end{eqnarray}  
The values of $I,J,K,...= 0,...,9,11$ parametrize the full eleven-dimensional space
$M_{11}$, while $\bar I, \bar J, \bar K,...=0,...,9$ are used for the ten-dimensional
hyperplanes, $M^{(i)}_{10}$, $i = 1,2$, orthogonal to the orbifold.  
The $F^{(i)}_{\bar I \bar J}$ are the two $E_8$ gauge field strengths and  
$C_{IJK}$ is the 3-form with field strength given by $G_{IJKL} = 24 \d_{[I}C_{JKL]}$.
In order for this theory to be supersymmetric and anomaly free the Bianchi identity for $G$
must pick up the following correction,
\be\label{eq:bianchi}
(dG)_{11 \bar I \bar J \bar K \bar L} = 
- \frac{1}{2\sqrt 2 \pi} \left( \frac{\kappa}{4\pi} \right)^{2/3} \left\{ J^{(1)}
\delta(x^{11}) + J^{(2)}\delta(x^{11} - \pi \rho) \right\}_{\bar I \bar J \bar K \bar L}
\ee
where the sources are
\be\label{eq:source}
J^{(i)} = tr F^{(i)} \wedge F^{(i)} - \frac{1}{2} tr R \wedge R
\,.
\ee

Now, we search for solutions to the above theory which preserve four of 
the thirty-two supercharges and, when compactified, lead to four dimensional,
$N = 1$ supergravities.  To begin, consider the manifold 
${\mathcal{M}} = I\!\! R^4 \times X \times S^1/Z_2$, where $I\!\! R^4$ is 
four-dimensional Minkowski space and $X$ is a Calabi-Yau three-fold.  Upon compactification
onto $X$, we are left with a five-dimensional effective
spacetime consisting of two copies of $ I\!\! R^4$, one at each of the orbifold fixed points,
and the orbifold itself (see fig. (5.2)).  On each of the $ I\!\! R^4$ planes there is a gauge
group $H^{(i)}$, $i = 1,2$, and $N=1$.  

\vspace{.75cm}
\hglue 2.5cm
\psfig{figure=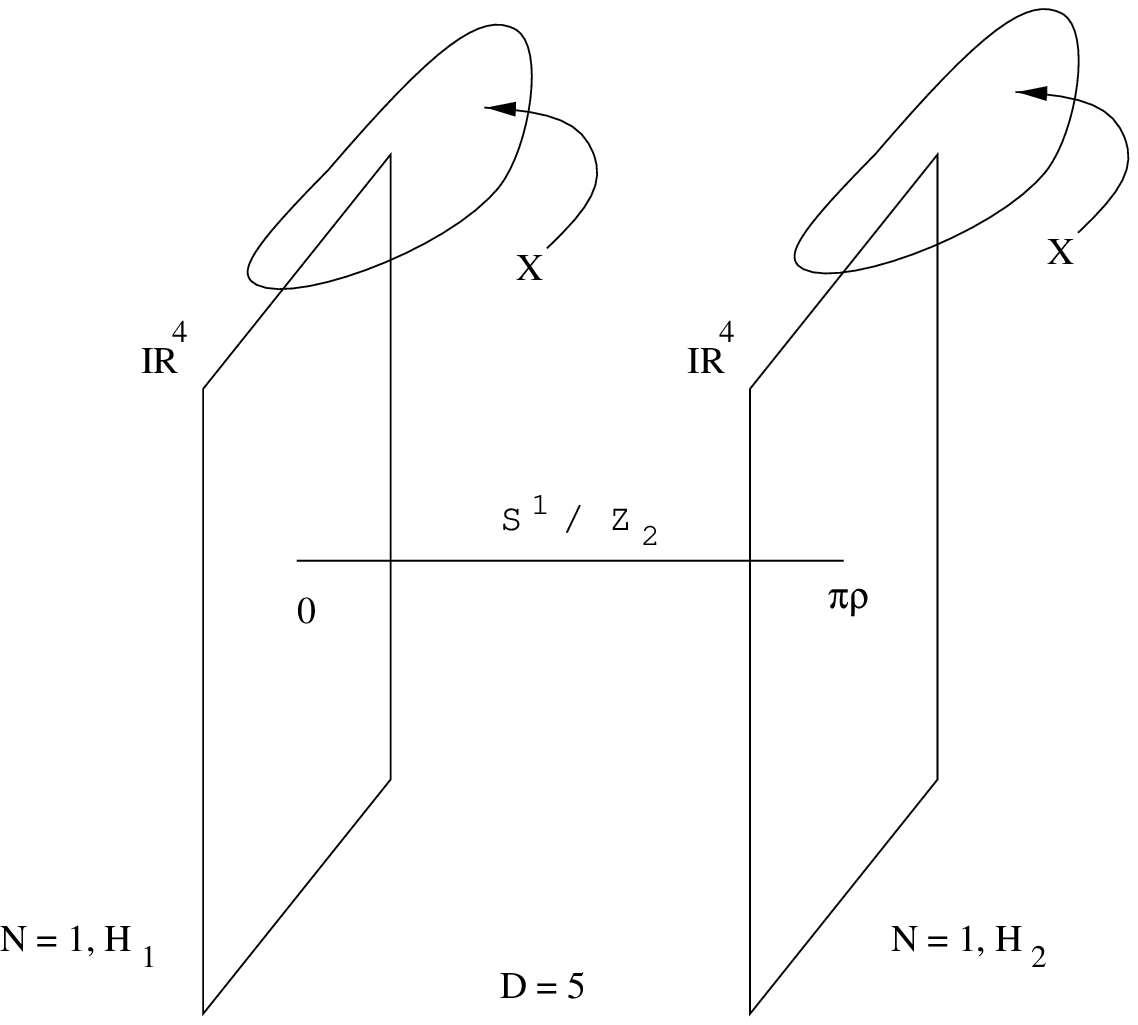,height=7.5cm,angle=-90}
\begin{quote}
\scriptsize Figure 5.2:
The LOW scenario.  The manifold is given by 
${\mathcal{M}} = I\!\! R^4 \times X \times S^1/Z_2$ 
where the Ho\v{r}ava-Witten theory is compactified on a smooth
Calabi-Yau three fold $X$.  
Compactification results in a five-dimensional
effective theory.
\end{quote}
In the next section we construct the
five-dimensional effective theory. 

\subsection{Five-Dimensional Effective Theory}\label{fived}

As we have discussed, according to the model presented above, there is an epoch
when the universe appears to be five dimensional.  Hence, it is only natural
to try to find the action for this five-dimensional effective theory.  Let us
identify the fields in the five-dimensional bulk.  First, there is the gravity
multiplet $(g_{\a\b}, {\mathcal{A}}_\a, \psi^i_\a)$, where $g_{\a\b}$ is the graviton,
${\mathcal{A}}_\a$ is a five-dimensional
vector field, and the $\psi^i_\a$ are the gravitini. The indices $\a,\b = 0,...,3,11$ and
$i = 1,2$.  There is also the universal hypermultiplet $q \equiv (V,\sigma, \xi,\bar \xi,\zeta^i)$.
Here $V$ is a modulus field associated with the volume of the Calabi-Yau space,
$\xi$ is a complex scalar zero mode, $\sigma$ is a scalar resulting
from the dualization of the three-form $C_{\a\b\g}$, and the $\zeta^i$ are the
hypermultiplet fermions.

It is now possible, using the action (\ref{eq:action}) to construct the five-dimensional
effective action of Ho\v{r}ava-Witten theory,
\be\label{eq:action5}
S_5 = S_{grav} + S_{hyper} + S_{bound}
\,,
\ee
where,
\be\label{eq:sgrav}
S_{grav}  =  - \frac{v}{2 \kappa^2}\int_{M_5} \sqrt{- g} \, \bigg[ R + \frac{3}{2}
        ({\mathcal{F}}_{\a \b})^2 
         + \frac{1}{\sqrt 2} \epsilon^{\a \b \g \delta \epsilon} {\mathcal{A}}_\a 
        {\mathcal{F}}_{\b \g} {\mathcal{F}}_{\delta \epsilon} \bigg]
\,,
\ee
\be\label{eq:shyper}
S_{hyper} = - \frac{v}{2 \kappa^2}\int_{M_5} \sqrt{- g} \, \bigg[ 4 \, h_{\m\n}\, \nabla_\a q^\m \,
        \nabla^\a q^\n + \frac{\a^2}{3} V^{-2} \bigg]
\,,
\ee
\begin{eqnarray}\label{eq:sbound}
S_{bound} & = & - \frac{v}{2 \kappa^2} \bigg[ \mp 2 \sqrt 2 \, 
        \sum_{i=1}^2 \int_{M_4^{(i)}} \sqrt{- g} \,
        \a \, V^{-1} \bigg] \nonumber \\
        & &- \frac{v}{8 \pi \kappa^2} \left( \frac{\kappa}{4\pi} \right)^{2/3} \,
        \sum_{i=1}^2 \int_{M_4^{(i)}} \sqrt{- g} \, V \,tr (F_{\m\n}^{(i)})^2
\,.
\end{eqnarray}  
In the above, $v$ is a constant that relates the five-dimensional Newton constant, $\kappa_5$,
with the eleven-dimensional Newton constant, $\kappa$, via $\kappa_5^2 = \kappa^2/v$.
The metric $h_{\m\n}$ is the flat space metric and $\a$ is a constant.  Higher-derivative 
terms have been dropped and this action provides us with a minimal $N=1$ supergravity 
theory in the five-dimensional bulk.

This theory admits a three-brane domain wall solution
with a world-volume lying in the four uncompactified dimensions \cite{ref:lukaset1998}.  
In fact, a pair of domain walls
is the vacuum solution of the five-dimensional theory which provides us with a 
background for reduction to a $d=4$, $N=1$ effective theory.  This solution will be
the topic of the next section.

\subsection{Three-Brane Solution}\label{branes}

In order to find a pair of three-branes solution we should start with an ansatz for
the five-dimensional metric of the form
\begin{eqnarray}\label{eq:metric}
ds_5^2 & = & a(y)^2 \, dx^\m \, dx^\n \, \eta_{\m\n} + b(y)^2 dy^2 \\
V & = & V(y) \nonumber
\,,
\end{eqnarray}
where $y = x^{11}$.
By using the equations of motion derived from the action (\ref{eq:action5}) we
find
\begin{eqnarray}\label{eq:solutions}
a & = & a_0 H^{1/2} \\
b & = & b_0 H^2 \nonumber \\
V & = & b_0 H^3 \nonumber
\,,
\end{eqnarray}
where $H \equiv \frac{\sqrt 2}{3} \alpha_0 |y| + c_0$, and $a_0,\, b_0$ and $c_0$ are
all constants.  Using the equations of motion derived by varying the
action with respect to $g_{\m\n}$ of (\ref{eq:metric}), we arrive
at a differential equation which leads to
\be\label{eq:julia}
         \d_y^2 H = \frac{2 \sqrt 2}{3} \, \a_0 \, \left( \delta(y)
                - \delta(y - \pi \rho) \right)
\,.
\ee
A detailed derivation of this equation is discussed in 
\cite{ref:lukaset1998}.  Clearly, (\ref{eq:julia})
represents two parallel three-branes located at the orbifold planes, as 
in Fig.(5.2).  This solves
the five-dimensional theory exactly and preserves half of the supersymmetries, with
low-energy gauge and matter fields carried on the branes.  This prompts us 
to find realistic cosmological models from the above scenario where the universe lives
on the world-volume of a three-brane.

\subsection{Cosmological Domain-Wall Solution}\label{cosmolowall}

In order to construct a dynamical, cosmological solution,
the solutions in (\ref{eq:solutions}) are made to be
functions of time $\t$, as well as the eleventh dimension $y$,
\begin{eqnarray}\label{eq:newmet}
ds_5^2  & = & - N(\t,y)d\t^2 + a(\t,y)^2 \, dx^m \, dx^n \, \eta_{mn} + b(\t,y)^2 dy^2 \\
V & = & V(\t, y) \nonumber
\,.
\end{eqnarray}
Here we have introduced a lapse function $N(\t,y)$.  Because this ansatz leads
to a very complicated set of non-linear equations we will seek a solution based on the
separation of variables.  Note, there is no \it a priori \rm reason to believe
that such a solution exists, but we will see that one does.  Separating the variables
$\t$ and $y$,
\begin{eqnarray}\label{eq:separate}
N(\t,y) & = & n(\t) a(y) \\
a(\t,y) & = & \a(\t) a(y) \nonumber \\
b(\t,y) & = & \b(\t) b(y) \nonumber \\
V(\t, y) & = & \g(\t) V(y) \nonumber 
\,.
\end{eqnarray}
Since this article is intended only as an elementary review we will not repeat
the details involved in solving the above system.  For our purposes it suffices
to say that the equations take on a particularly simple form when $\b = \g$ and
with the gauge choice of $n = const.$.  In this gauge, $\t$ becomes proportional
to the comoving time $t$, since $dt = n(\t) d\t$.  A solution exists
such that
\begin{eqnarray}\label{eq:expwall}
\a & = & A \, |t-t_0|^p \\
\b & = & \b = B \, |t-t_0|^q \nonumber
\,,
\end{eqnarray}
where 
\begin{eqnarray}\label{eq:pandq}
p & = & \frac{3}{11} (1 \mp \frac{4}{3\sqrt 3}) \\
q & = & \frac{2}{11} (1 \pm {2\sqrt 3}) \nonumber
\,,
\end{eqnarray}
and $A,B$ and $t_0$ are arbitrary constants.  This is the desired cosmological
solution.  The $y$-dependence is identical to the domain wall solution (\ref{eq:julia})
and the scale factors evolve with $t$ according to (\ref{eq:expwall}).  The domain
wall pair remain rigid, while their sizes and the separation between the walls
change.  In particular, $\a$ determines the size of the domain-wall world-volume while
$\b$ gives the separation of the two walls.  In other words, $\a$ determines the size
of the three-dimensional universe, while $\b$ gives the size of the orbifold.
Furthermore, the $d=4$ world-volume of the three-brane universe exhibits $N=1$ SUSY 
(of course SUSY is broken in the dyanamical solution) and a particular solution 
exists for which the domain wall world-volume expands in a 
FRW-like manner while the orbifold radius contracts.  

Although the above model provides an intriguing use of M-theory in an attempt to 
answer questions about early universe cosmology there are still many problems to be worked out.
Foremost, these are vacuum solutions, devoid of matter and radiation.  There
is no reason to think that, of all the solutions, the one which matches our universe
(expanding domain-wall, shrinking orbifold) should be preferred over any other.  This problem
is typical of many cosmological models, however.  The Calabi-Yau
(six-dimensional) three-fold is chosen by hand in order to give four noncompact dimensions.
Hence, the dimensionality problem mentioned in Section \ref{dimen} is still present in this model.
Stabilization of moduli fields, including the dilaton has recently been addressed in \cite{ref:hueyet2000}.
There are no cosmological constants in the model.  There is also no natural mechanism supplied for SUSY breaking on the
domain wall, and currently no discussion of inflationary dynamics.  For more on heterotic
M-theory and cosmology see, \cite{ref:horavaet1996a}-\cite{ref:donagietb2000}.

\section{Large Extra Dimensions}\label{led}

This section provides a brief discussion of scenarios involving
large extra dimensions, focusing primarily on the models of Randall and Sundrum
(RSI and RSII)\footnote{The distinction between RSI and RSII models will be clarified below.}
\cite{ref:randallet1999, ref:randallsu1999}.  The RSI model is similar in 
many respects to that of the Lukas, Ovrut and Waldram scenario discussed in
section \ref{heterotic}, although its motivation is quite different.  In the LOW
construction the motivation was to construct a cosmology out of the fundamental theory
of everything.  In the RSI model the motivation is to construct a cosmology
in which the Hierarchy problem of the Standard Model (SM) is 
solved in a natural way.  Some earlier proposals involving large extra dimensions
include \cite{ref:dvaliet1999b}-\cite{ref:akama2000j}.  Also see the extensive set
of references in \cite{ref:pavsicet2000}.

\subsection{Motivation and the Hierarchy Problem}\label{hier}

There is a hierarchy problem in the Standard Model because we have no way
of explaining why the scales of particle physics are so different from those of gravity.
Many attempts to solve the hierarchy problem using extra dimensions have been made
before, see for example \cite{ref:dvaliet1999a} and \cite{ref:dvaliet1999b}. 
If spacetime is fundamentally $(4+n)$-dimensional then the physical Planck mass
\be\label{eq:pmass}
M_{pl}^{(4)} \simeq 2 \times 10^{18} \GeV
\,,
\ee
is actually dependent on the fundamental $(4+n)$-dimensional Planck mass $M_{pl}$ 
and on the geometry of the extra dimensions according to
\be\label{eq:fmass}
{M_{pl}^{(4)}}^2 = M_{pl}^{n+2}\, V_n
\,,
\ee
Here $V_n$ is the volume of the $n$ compact
extra dimensions.  Because we have not detected any extra dimensions experimentally,
the compactification scale $\m_c \sim 1/V_n^{1/n}$ would have to be much smaller
than the weak scale, and the particles and forces of the SM (except for gravity)
must be confined to the four-dimensional world-volume of a three-brane 
(See \fig{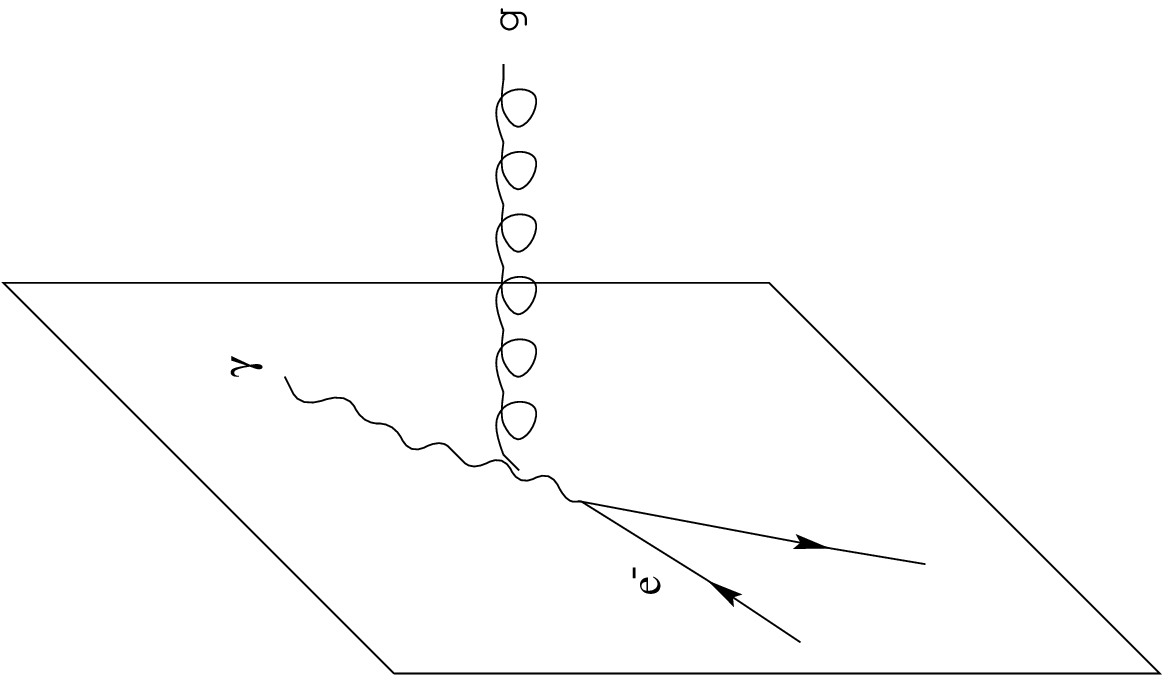}). 

\vspace{.75cm}
\hglue 5.5cm
\psfig{figure=conf.eps,height=7.5cm,angle=-90}\label{conf.eps}
\begin{quote}
\scriptsize Figure \ref{conf.eps}:  In the RS model the fields of the
Standard Model (with the exception of gravity) are confined to the
three-brane world-volume while gravity is allowed to propagate in the bulk.
\end{quote}

We see from (\ref{eq:fmass}) that by taking $V_n$ to be large enough it is possible to
eliminate the
hierarchy between the weak scale $v$ and the Planck scale.  Unfortunately, 
in this procedure a new hierarchy has been introduced, namely the one between $\m_c$ and $v$.
Randall and Sundrum proposed the following:
We assume that the particles and forces
of the SM with the exception of gravity are confined to a four-dimensional 
subspace of the $(4+n)$-dimensional spacetime.  This subspace is identified with the world-volume
of a three-brane and an ansatz for the metric is made.  Randall and Sundrum's proposal is
that the metric is not factorizable, but
the four-dimensional metric is multiplied by a ``warp" factor that is exponentially
dependent upon the radius of the bulk, fifth dimension.  The metric ansatz is
\be\label{eq:rsmetric}
ds^2 = e^{-2 k r_c \varphi} \, \eta_{\m\n} \, dx^\m dx^\n + r_c^2\,d\varphi^2
\,,
\ee
where $k$ is a scale of order the Planck scale, $\eta_{\m\n}$ is the four-dimensional
Minkowski metric and $0 \le \varphi \le \pi$ is the coordinate for the
extra dimension.  Randall and Sundrum have shown that this metric solves the
Einstein equations and represents two three-branes with appropriate cosmological constant
terms separated by a fifth dimension.  The above scenario, in addition to being
able to solve the hierarchy problem (see section \ref{fourd}), provides distinctive experimental signatures.
Coupling of an individual
Kaluza-Klein (KK) excitation to matter or to other gravitational modes is set by the weak and not
the Planck scale.  There are no light KK modes because the excitation scale is
of the order a TeV.  Hence, it should be possible 
to detect such excitations at accelerators (such as the LHC).  
The KK modes are observable as spin $2$ excitations that can be reconstructed from
their decay products.  For experimental signatures of KK modes within large extra
dimensions see e.g. \cite{ref:buettneret2000, ref:cheung2000, ref:rizzo2000}.

\subsection{Randall-Sundrum I}\label{rsi}

The basic setup for the RSI model is depicted in \fig{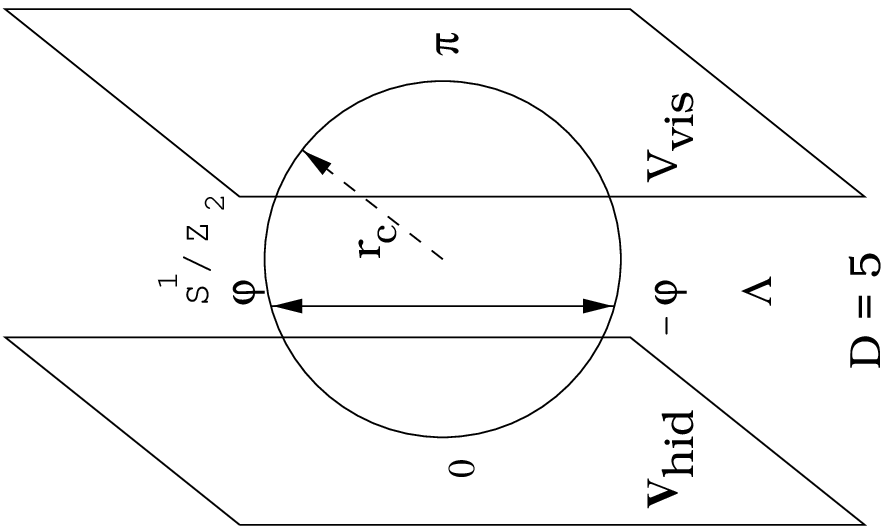}.  
The angular coordinate $\varphi$ parameterizes
the fifth dimension and ranges from $-\pi$ to $\pi$.  The fifth dimension is taken as
the orbifold $S^1/Z_2$ where there is the identification of $(x,\varphi)$ with
$(x,-\varphi)$.  The orbifold fixed points are at $\varphi = 0,\pi$ and correspond
with the locations of the three-brane boundaries of the five-dimensional spacetime.
Note the similarities of this model with the LOW model of Section \ref{heterotic}.
One difference is that we are now considering nonzero vacuum energy densities on
both the visible and the hidden brane and in the bulk.  

\vspace{.75cm}
\hglue 5cm
\psfig{figure=rsbrns.eps,height=7.5cm,angle=-90}\label{rsbrns.eps}
\begin{quote}
\scriptsize Figure \ref{rsbrns.eps}:  The Randall-Sundrum scenario.  The fifth dimension is
compactified onto the orbifold $S^1/Z_2$.
\end{quote}

The action describing the scenario is
\be\label{eq:rsaction}
S = S_{grav} + S_{vis} + S_{hid}
\ee
where
\begin{eqnarray}\label{eq:rss}
S_{grav} & = & \int d^4 x \int_{-\pi}^\pi d\varphi \, \sqrt{- G} \, \big( - \Lambda
        + 2M^3 R \big) \nonumber \\
S_{vis} & = & \int d^4 x \, \sqrt{- g_{vis}} \, \big( {\mathcal{L}}_{vis} + V_{vis} \big) 
                        \nonumber \\
S_{hid} & = & \int d^4 x \, \sqrt{- g_{hid}} \, \big({\mathcal{L}}_{hid} + V_{hid} \big) 
\,.
\end{eqnarray}
Here, $M$ is the Planck mass, $R$ is the Ricci scalar,
$g_{vis}$ and $g_{hid}$ are the four-dimensional metrics on the visible and
hidden sectors respectively and $V_{vis}$, $\Lambda$ and $V_{hid}$ are the
cosmological constant terms in the visible, bulk and hidden sectors.
The specific form for the three-brane Lagrangians is not relevant for
finding the classical five-dimensional, ground state metric.
The five-dimensional Einstein equations for the above action are
\begin{eqnarray}\label{eq:5deinst}
\sqrt{- G} \; \Big(R_{MN} - \frac{1}{2} G_{MN} R \Big) & = & -\frac{1}{4M^3}
        \Big[ \Lambda \sqrt{- G} \; G_{MN} \nonumber \\
& & + V_{vis} \; \sqrt{- g_{vis}} \; g_{\m\n}^{vis} \; 
        \delta^\m_M \; \delta^\n_N \; \delta(\varphi - \pi) \nonumber \\
& & + V_{hid} \; \sqrt{- g_{hid}} \; g_{\m\n}^{hid} \; \delta^\m_M \; \delta^\n_N \; 
        \delta(\varphi)
\,.
\end{eqnarray}
We now assume that a solution exists which has four-dimensional Poincar\'{e} invariance
in the $x^\m$ directions.  A five-dimensional ansatz which obeys the above requirements is
\be\label{eq:5dans}
ds^2 = e^{- 2\s(\varphi)} \, \eta_{\m\n} dx^\m dx^\n + r_c^2 \, d\varphi^2
\,.
\ee
Substituting this ansatz into (\ref{eq:5deinst}) reduces the Einstein equations to
\begin{eqnarray}
\frac{6 \s'^2}{r_c^2} & = & \frac{-\Lambda}{4M^3} \,, \label{eq:einst1} \\
\frac{3 \s''}{r_c^2} & = & \frac{V_{hid}}{4 M^3 r_c}\delta(\varphi) 
        + \frac{V_{vis}}{4 M^3 r_c}\delta(\varphi - \pi) \label{eq:einst2}
\,.
\end{eqnarray}
Solving (\ref{eq:einst1}) consistently with orbifold symmetry 
$\varphi \rightarrow -\varphi$,  we find
\be\label{eq:sigma}
\s = r_c \, |\varphi| \, \sqrt{\frac{- \Lambda}{24 M^3}}
\,,
\ee
which makes sense if $\Lambda < 0$.  With this choice, the spacetime in the bulk of
the theory is a slice of an $AdS_5$ manifold.   Also, to solve
(\ref{eq:einst2}) we should take 
\begin{eqnarray}\label{eq:costerms}
V_{hid}  =  -V_{vis} & = & 24M^3k \nonumber \\
\Lambda & = & - 24 M^3k^2
\,.
\end{eqnarray}
Note that the boundary and bulk cosmological terms are dependent upon the
single scale factor $k$, and that the relations between them are required in
order to get four-dimensional Poincar\'{e} invariance.

Further connections with the LOW scenario of Section \ref{heterotic} are
now visible.
The exact same relations given in (\ref{eq:costerms}) arise in the 
five-dimensional Ho\v{r}ava-Witten effective theory if one identifies the
expectation values of the background three-form field as cosmological terms \cite{ref:lukasetb1999}.

We want the bulk curvature to be small compared to the higher dimensional
Planck scale in order to trust the solution and thus, we assume $k<M$.
The bulk metric solution is therefore,
\be\label{eq:finalm}
ds^2 = e^{- 2kr_c |(\varphi)|} \, 
        \eta_{\m\n} dx^\m dx^\n + r_c^2 \, d\varphi^2
\,.
\ee
Since $r_c$ is small but still larger than $1/k$, the fifth dimension cannot be
experimentally observed in present or future gravity experiments.  This prompts
us to search for a four-dimensional effective theory.

\subsubsection{Four-Dimensional Effective Theory}\label{fourd}

In our four-dimensional effective description we wish to find the parameters of
this low-energy theory (e.g. $M_{pl}^{(4)}$ and mass parameters of the four-dimensional
fields ) in terms of the five-dimensional, fundamental scales,
$M$, $k$ and $r_c$.  In order to find the four-dimensional
theory one identifies massless gravitational fluctuations about the classical
solution \ref{eq:finalm} which correspond to the gravitational fields for the effective theory.  These
are the zero modes of the classical solution.  The metric of the four-dimensional
effective theory is of the form
\be\label{eq:effm}
\bar g_{\m\n}(x) \equiv \eta_{\m\n} + \bar h_{\m\n}(x)
\,,
\ee
which is locally Minkowski.  Here, $ \bar h_{\m\n}(x)$ represents the tensor fluctuations about
Minkowski space and gives the physical graviton of the four-dimensional effective theory.
By substituting the metric (\ref{eq:effm}) for $\eta_{\m\n}$ in (\ref{eq:finalm}) and
then using the result in the action (\ref{eq:rss}) the curvature term becomes 
\be\label{eq:stuff}
S_{eff} \propto \int d^4x \int_{-\pi}^\pi d\varphi \, 2M^3 r_c 
        e^{-2kr_c|\varphi|}\,\sqrt{- \bar g}\,\bar R
\,,
\ee
where $\bar R$ is the four-dimensional Ricci scalar made out of $\bar g_{\m\n}(x)$.
We focus on the curvature term so that we may derive the scale of the
gravitational interactions.
The effective fields depend only on $x$, and hence it is possible to perform the integration
over $\varphi$ explicitly, obtaining
the four-dimensional
effective theory \cite{ref:randallet1999}.  Using the result one may derive an expression
for the four-dimensional Planck mass in terms of the fundamental, five-dimensional Planck mass
\be\label{eq:mofm}
{M_{pl}^{(4)}}^2 = M^3\, r_c \, \int_{-\pi}^\pi d\varphi \, e^{-2kr_c|\varphi|}
        = \frac{M^3}{k}[1 - e^{-2kr_c\pi}]
\,.
\ee
Notice that ${M_{pl}^{(4)}}$ depends only weakly on $r_c$ in the large $kr_c$ limit.

From the fact that
$g_{\m\n}^{vis}(x^\m) \equiv G_{\m\n}(x^\m,\varphi = \pi)$ and
$g_{\m\n}^{hid}(x^\m) \equiv G_{\m\n}(x^\m,\varphi = 0)$
we find,
\be\label{eq:ghid}
\bar g_{\m\n} = g_{\m\n}^{hid}
\,,
\ee
but
\be\label{eq:gvis}
\bar g_{\m\n} = g_{\m\n}^{vis} e^{2kr_c \pi}
\,.
\ee
It is now possible to find the matter field Lagrangian of the theory.
With proper normalization of the fields one can determine physical masses.  Let us 
consider the example of a fundamental Higgs field.  The action is
\be\label{eq:avis}
S_{vis} \simeq \int d^4 x \sqrt{-g_{vis}} \; \left( g_{vis}^{\m\n} \, D_\m H^{\dagger}
        D_\n H - \lambda \left(|H|^2 - v_0^2 \right)^2 \right)
\,,
\ee
which contains only one mass parameter $v_0$.  Using (\ref{eq:gvis}) the action becomes
\be\label{eq:avist}
S_{eff} \simeq \int d^4 x \sqrt{- \bar g} e^{- 4kr_c \pi} \; 
        \left( \bar g^{\m\n} e^{2kr_c \pi} \, D_\m H^{\dagger}
        D_\n H - \lambda \left(|H|^2 - v_0^2 \right)^2 \right)
\,,
\ee
and after wavefunction renormalization, $H \rightarrow e^{kr_c \pi} H$, we have
\be\label{eq:afin}
S_{eff} \simeq \int d^4 x \sqrt{- \bar g} \; \left( \bar g^{\m\n} \, D_\m H^{\dagger}
        D_\n H - \lambda \left(|H|^2 - e^{- 2kr_c \pi}v_0^2 \right)^2 \right)
\,.
\ee
This result is completely general.  The physical mass scales are set by a symmetry-breaking
scale,
\be\label{eq:v0stuff}
v \equiv e^{- kr_c \pi}v_0
\,,
\ee
and hence any mass parameter $m_0$  on the visible three-brane is related to the fundamental, 
higher-dimensional mass via
\be\label{eq:masses}
m \equiv e^{- kr_c \pi}m_0
\,.
\ee
Note that
if $ e^{ kr_c \pi} \sim 10^{15}$, TeV scale physical masses are produced
from fundamental mass parameters near the Planck scale, $10^{19} \, \GeV$. 
Therefore, there are no large hierarchies if $kr_c \approx 50$.  

\subsection{Randall-Sundrum II}\label{rsii}

In the RSI scenario described in the last section our universe was identified with the
negative tension brane while the brane in the hidden sector had positive tension 
(Eq. (\ref{eq:costerms})).  In this model
it was shown that the hierarchy problem may be solved.  Unfortunately, there are
several problems with the idea that the universe we live in can be
a negative tension brane.  For one, the energy density of matter on such
a brane would be negative and gravity repulsive 
\cite{ref:csa'kiet1999, ref:clineet1999, ref:shiromizuet2000}.
Life is more comfortable on a positive tension brane since the D-branes 
which arise as fundamental objects in string theories are all positive
tension objects and the localization of matter and gauge fields on positive
tension branes is well understood within the context of string theory.
  
For the above reasons Randall and Sundrum suggested a second
scenario (RSII) in which our universe is the positive 
tension brane and the hidden brane has negative tension \cite{ref:randallsu1999}.
In this case the boundary and bulk cosmological constants are related by
\begin{eqnarray}\label{eq:rsii}
V_{vis}  =  -V_{hid} & = & 24M^3k \nonumber \\
\Lambda & = & - 24 M^3k^2
\,,
\end{eqnarray}
as opposed to the realation in RSI, Eq. (\ref{eq:costerms}).

As we will see, in this refined scenario it is possible to reproduce Newtonian
gravity and other four-dimensional general relativistic predictions at low
energy and long distance on the visible brane.  Note that in the solution
to the hierarchy problem in RSI the wave function for the massless graviton
is greatest on the hidden brane, whereas RSII has the graviton bound to the visible brane.
To see this, consider the wave equation for small gravitational fluctuations,
\be\label{eq:gravi}
\left(\d_\m \d^\m - \d_i \d^i + V(z_i) \right) \hat{h} (x^\m,z_i) = 0
\,.
\ee
This has a non-trivial potential term $V$ resulting from the curvature,  
$\m$ runs from $0$ to $3$ and $i$ labels the extra dimensions.  
It is possible to write $\hat h$ as a superposition of modes $\hat h = e^{i p \cdot x} \hat{\psi}(z)$
where $\hat\psi$ is an eigenmode of the equation
\be\label{eq:kk}
\left(-\d_i \d^i + V(z) \right) \hat{\psi}(z) = - m^2 \hat{\psi}(z)
\,,
\ee
in the extra dimensions and $p^2 = m^2$.  Hence, the higher-dimensional gravitational
fluctuations are Kaluza-Klein reduced in terms of four-dimensional KK states with
mass $m^2$ given by the eigenvalues of (\ref{eq:kk}).  The zero mode that is also
a normalizable state in the spectrum of Eq. (\ref{eq:kk}) is the wave function
associated with the four-dimensional graviton.  This state is a bound state
whose wave function falls off rapidly away from the 3-brane.  Such behavior 
corresponds to a 3-brane acting as a positive tension source on 
the right hand side of Einstein's equations.

The procedure of RSII is to decompactify the orbifold of RSI (i.e. consider
$r_c \rightarrow \infty$) taking the hidden, negative tension brane off to infinity.
In doing this, one obtains an effective four-dimensional theory of gravity where
the setup is a single three-brane with positive tension embedded in a five-dimensional
bulk spacetime.  On this brane one can compute an effective nonrelativistic
gravitational potential between two particles of masses $m_1$ and $m_2$ which is
generated by exchange of the zero-mode and continuum Kaluza-Klein mode propagators.
The potential behaves as
\be\label{eq:rsiipot}
V(r) = G_N \frac{m_1 m_2}{r} \left( 1 + \frac{1}{r^2}{k^2} \right)
\,.
\ee
Here the leading term is the usual Newtonian potential and is due to the bound
state mode.  The KK modes generate the $1/r^3$ correction term which is 
heavily suppressed for $k$ of order the fundamental Planck scale and 
$r$ of the size tested with gravity.
The propagators calculated in \cite{ref:randallsu1999} are relativistic and
hence, going beyond the nonrelativistic approximation one recovers all the 
proper relativistic corrections with negligible corrections from the continuum modes.

Let us compare the RSI and RSII models.  In RSI, the solution to 
the hierarchy problem requires that we are living on a negative tension
brane.  The positive tension brane has no such suppression of its masses
and is therefore often referred to as the ``Planck" brane, which is
hidden from the visible brane.  Serious arguments against this scenario
are that the negative tension ``TeV" brane seems physically unacceptable.

In RSII, the visible brane is taken as the positive tension brane
while the TeV brane is sent off to infinity.  In this model 
the proper Newtonian gravity is manifest on the visible brane, but the
hierarchy problem is not addressed.

Although more successful as a potential physical model of our universe than its 
predecessor RSI, RSII seems to lack the elegant solution to the hierarchy problem made
possible by considering the universe as a negative tension brane.  Recent
work however suggests that by including quantum effects (analogous to the 
Casimir effect) it is possible to solve the hierarchy problem on the visible
brane having either positive or negative tension \cite{ref:mukohyama2000}.  If the
Casimir energy is negative and one accepts a degree of fine tuning of the tension
on the hidden brane it is possible to obtain a large enough warp factor to
explain the hierarchy on the visible brane having either positive or negative tension.
Further work on this scenario is needed however including a study of the
stability of this model against perturbations.

\subsection{RS and Brane World Cosmology}

The next obvious step is to consider the cosmologies of the RS model discussed
above.  There has been an extensive amount of work
done in these areas and the reader is invited to examine the references at the
end of the review related to Randall-Sundrum and ``brane world" cosmologies 
\cite{ref:kimlee2000} - \cite{ref:goldbergeret2000} for 
a comprehensive study.  Due to the vast number of cosmological models discussed
in the literature we will review only the basics and focus on the problems of brane world
cosmologies while mentioning potential resolutions and future work, referencing various relevant authors.
Much of the discussion in this section closely parallels the excellent review of 
J. Cline \cite{ref:cline1999}.

We begin by considering the cosmological expansion of 
3-brane universes in a 5-dimensional
bulk with a cosmological constant as discussed by 
Bin\'{e}truy, Deffayet, Ellwanger and Langlois (BDEL) \cite{ref:binetruyetb2000}.
Note that in an earlier work \cite{ref:binetruyeta2000}, BDL considered
the solutions to Einstein's equations in five dimensions with an $S_1/Z_2$ orbifold
and matter included on the two branes but with no cosmological constants on the
branes or in the bulk.  They found that the Hubble
expansion rate of the visible brane was related to the energy density of the brane
quadratically opposed to the standard Friedmann equation, $H^2 \propto \r$.  We will
show this explicitly below.
The altered expansion rate proved to be incompatible with nucleosynthesis constraints.

When the analysis was applied to the RSII scenario one does in fact reproduce
the ordinary FRW universe on the positive tension, Planck brane \cite{ref:csa'kiet1999, ref:clineet1999}.  
Note however, that in the RSII scenario on the negative tension brane where the
hierarchy problem is solved the Friedmann equation has a critical sign difference.

In the BDEL model the authors consider five-dimensional spacetime metrics of the form
\be\label{eq:5dbdl}
ds^2 = \tilde{g}_{AB}\,dx^A \, dx^B = g_{\m\n}\,dx^\m \, dx^\n + b^2 dy^2 
\,
\ee
where $y$ is the coordinate associated with the fifth dimension.  The visible universe
is taken to be the hypersurface at $y=0$.  The metric is taken to be
\be\label{eq:bdlmet1}
ds^2 = -n^2(\t,y) d\t^2 + a^2(\t,y) g_{ij}dx^i dx^j + b^2(\t,y) dy^2
\,,
\ee
where $\g_{ij}$ is a maximally symmetric three-dimensional metric ($k=-1,0,1$ will parametrize 
the spatial curvature), and $n$ is a lapse function.

The five-dimensional Einstein equations have the usual form
\be\label{eq:einst5}
\tilde{G}_{AB} \equiv \tilde{R}_{AB} - \frac{1}{2} \tilde{R} \tilde{g}_{AB} = \k^2 \tilde{T}_{AB}
\,,
\ee
where $\k$ is related to the five-dimensional Newton's constant $G_{(5)}$ and the
five-dimensional reduce Planck mass $M_{(5)}$ by
\be\label{eq:kapgm}
\k^2 = 8\pi G_{(5)} = M^{-3}_{(5)}
\,.
\ee
Using the ansatz (\ref{eq:bdlmet1}) one finds the non-vanishing components of the
Einstein tensor to be
\begin{eqnarray}
{\tilde G}_{00} &=& 3\left\{ \fda \left( \fda+ \fdb \right) - \frac{n^2}{b^2} 
\left(\fppa + \fpa \left( \fpa - \fpb \right) \right) + k \frac{n^2}{a^2} \right\}, 
\label{ein00} \\
 {\tilde G}_{\ii\jj} &=& 
\frac{a^2}{b^2} \gamma_{ij}\left\{\fpa
\left(\fpa+2\fpn\right)-\fpb\left(\fpn+2\fpa\right)
+2\fppa+\fppn\right\} 
\nonumber \\
& &+\frac{a^2}{n^2} \gamma_{ij} \left\{ \fda \left(-\fda+2\fdn\right)-2\fdda
+ \fdb \left(-2\fda + \fdn \right) - \fddb \right\} -k \gamma_{ij},
\label{einij} \\
{\tilde G}_{05} &=&  3\left(\fpn \fda + \fpa \fdb - \frac{\dot{a}^{\prime}}{a}
 \right),
\label{ein05} \\
{\tilde G}_{55} &=& 3\left\{ \fpa \left(\fpa+\fpn \right) - \frac{b^2}{n^2} 
\left(\fda \left(\fda-\fdn \right) + \fdda\right) - k \frac{b^2}{a^2}\right\}.
\label{ein55} 
\end{eqnarray}
Here the prime indicates differentiation with respect to $y$ and dot 
indicates differentiation with respect to $\t$.

The energy-momentum tensor can be described in terms of the fields living on the
visible brane world-volume $T^A_B$ and the fields living in the bulk space 
(or on other branes) $\check{T}^A_B$. We have
\be\label{eq:momen}
T^A_B = \frac{\delta(y)}{b}\, diag(-\r,p,p,p,0)
\,,
\ee
where the energy density $\r$ and pressure $p$ are independent of the position in
the brane in order to recover a homogeneous cosmology on the brane.  The total
energy-momentum tensor is then
\be\label{eq:totmom}
\tilde{T}^A_B = T^A_B + \check{T}^A_B 
\,.
\ee
Note that
in reality the brane would have some thickness in the fifth dimension determined
by the fundamental scale of the underlying theory.  However, the presence
of the delta function in (\ref{eq:momen}) (the ``thin-brane" approximation) should
be valid when the energy scales are much smaller than the fundamental scale.
In what follows unless otherwise mentioned we will take $k=0$ as in \cite{ref:binetruyeta2000}.

From the Bianchi identity $\nabla_A \tilde{G}^A_B = 0$ and the Einstein equations 
(\ref{eq:einst5}) an equation of conservation is obtained,
\be\label{eq:conrs}
\dot\r + 3(p+\r)\frac{\dot a_0}{a_0} = 0
\,,
\ee
which matches the usual four-dimensional equation of energy density conservation in standard
cosmology.  Here $a_0$ is the value of $a$ on the brane.

To find a solution of Einstein's equations (\ref{eq:einst5}) in the vicinity of
the visible brane at $y=0$ one must deal with the delta function sources.  The details
may be found in \cite{ref:binetruyetb2000}.  From the $55$ component equation (\ref{ein55})
one finds a new Friedmann-like equation
\be\label{eq:rsfrw}
\frac{\dot a_0^2}{a_0^2} + \frac{\ddot a_0}{a_0} = 
        - \frac{\k_{(5)}^4}{36} \r (\r + 3p) - \frac{\k_{(5)}^2 \check{T}_{55}}{3b_0^2}
\,.
\ee
Immediately, and as mentioned above one sees the unusual quadratic dependence $H^2 \propto \r^2$.  

Note that if one allows for a cosmological constant in the bulk $\Lambda$ and 
extra energy densities on the Planck and TeV branes ($\r_P$ and $\r_T$, respectively),
in addition to the respective tensions $\s$ and $-\s$ one finds a Friedmann-like 
equation of the form \cite{ref:cline1999}\footnote{Now we have switched to the notation of \cite{ref:cline1999} but here
$\bar k$ is the $k$ introduced in our discussion of RSI, Section (\ref{rsi}).}
\be\label{eq:cfrw}
H^2 = \frac{(\s + \r_{p})^2}{36M^6} + \frac{\Lambda}{6M^3} = 
        \frac{(-\s + e^{-2\bar{k}b}\r_{T})^2}{36M^6} + \frac{\Lambda}{6M^3}
\,.
\ee
When the tension $\s$ is fine tuned to cancel the contribution from $\Lambda$ in the
limit $\r_i =0$ , it is possible to recover the correct FRW behavior $H\propto \sqrt \r$
at leading order in $\r$ \cite{ref:csa'kiet1999,ref:clineet1999,ref:cline1999}.  Interestingly, this
fine tuning is exactly that required by RS to obtain a static solution.  Unfortunately
while the cosmology on the Planck brane appeared normal,
the energy density on the TeV brane, where the hierarchy problem is solved, is negative
which is physically unacceptable as we mentioned above.

Exciting new developments have shown that when the radion is stabilized the
previously mentioned unconventional cosmologies in the RS model disappear \cite{ref:csa'kiet1999}.
By assuming that a 5-dimensional potential $U(b)$ is generated by some mechanism
(e.g. \cite{ref:goldbergeret2000}) in the 5-dimensional theory, the nonvanishing equations
of motion in the bulk (with cosmological constant $\Lambda$) reduce to 
\begin{eqnarray}\label{eq:solveme}
\tilde G_{00} &=& \k^2 n^2(\Lambda + U(b)) \,, \nonumber \\
\tilde G_{ii} &=& -\k^2 a^2(\Lambda + U(b)) \,, \nonumber \\
\tilde G_{55} &=& -\k^2 b^2(\Lambda + U(b) + U'(b)) \,.
\end{eqnarray}
Here $\tilde G_{AB}$ is given by (\ref{ein00})-(\ref{ein55}) with $k = 0$.
Let us introduce notation $m_0$ such that the static RS solution is recovered when
$V_p = -V_T = 6m_0/\k^2 $ and $\Lambda = - 6m_0^2/\k^2$.  We take the locations
of the Planck and TeV branes to be at $y = 0$ and $y=1/2$, respectively.
To simplify the solution of (\ref{eq:solveme}) the radion is assumed to be very heavy
and near its minimum $U \approx M_b^5 ((b-b_0)/b_0)^2$.  Here $b_0$ is the stabilized
value of $b$ and $M_b$ is proportional to the radion mass $m_{rad}$.  Now
one may perturb around the RS solution, with cosmological constants $\delta V_p$ and
$\delta V_T$ instead of matter densities.  Using the ansatz
\begin{eqnarray}\label{eq:radst}
a(t,y) &=&  e^{Ht-|y|m_0b_0}(1+\delta a(y)) \,,\nonumber \\
n(t,y) &=&  e^{-|y|m_0b_0}(1+\delta a(y))  \,, \nonumber \\
b&=&b_0 \,,
\end{eqnarray}
it is possible to derive the Friedmann equation
\be\label{eq:friedmo}
H^2 = \frac{\k^2 m_0}{3(1 - \Omega_0^2)} (\delta V_p + \delta V_T \Omega_0^4) 
\,,
\ee
where $\Omega_0 \equiv e^{-m_0 b_0/2}$.  Note that (\ref{eq:friedmo}) is the
standard Hubble law with correct normalization for the physically observed
energy density $\rho = \delta V_p + \delta V_T \Omega_0^4$.  

The constraint between the matter on the two branes was a consequence of trying
to find a static solution to the radion equations of motion without actually providing
a mechanism for stabilization.  Once such a mechanism is introduced the constraint
vanishes as described above, and the ordinary 4-dimensional FRW behavior is recovered at low temperatures
if the radion has a mass of order the weak scale.  It was suggested in \cite{ref:csa'kiet1999}
that matter on the hidden brane or in the bulk may be a dark matter candidate.

As we have already discussed above, it seems unlikely that the RSI scenario as presented 
in \cite{ref:randallet1999}
can provide a physically
realistic cosmological model as the energy density on the TeV brane is negative.  The RSII 
model, having non-compact extra dimension, has greater success as a cosmological model in
that it correctly reproduces the conventional cosmology on the visible brane (see e.g. \cite{ref:csa'kiet1999}).
Other variations of both RSI and RSII and alternative brane world models have also produced
correct cosmological behaviors of our universe.  The reader is referred to the review \cite{ref:cline1999} 
for a detailed summary of work on RS cosmology.  

Important problems and challenges which need to be explained in brane world scenarios
include the stabilization of the radius of the extra dimension and the radion field
\cite{ref:clineet2000} - \cite{ref:goldbergeret2000}, inflation \cite{ref:himemotoet2000}-\cite{ref:dvalitye1999},
incorporation into supergravity models \cite{ref:duffeti2000}-\cite{ref:cveticetij2000},
string theory and the AdS/CFT correspondence \cite{ref:kachru2000}-\cite{ref:hawkinget2000}.
In particular see the review \cite{ref:kachru2000} and the references therein.
For more on the cosmological constant and brane worlds see 
\cite{ref:kachruet2000,ref:verlinde2000k} and \cite{ref:li2000}-\cite{ref:hamedet2000}.
For early versions of brane world scenarios see \cite{ref:akama2000j}-\cite{ref:pavsicet2000}.
Experimental predictions are discussed in e.g. \cite{ref:buettneret2000}-\cite{ref:rizzo2000}.
Cosmologies of brane world scenarios are analyzed in \cite{ref:kimlee2000}-\cite{ref:kakushadzeet1999}.

\subsection{Supersymmetry}

We will have only a few comments in this section, as the work in this
area is still too new to review. There have been a number of attempts 
to include supersymmetry into the
RS and brane world scenarios \cite{ref:duffeti2000}-\cite{ref:cveticetij2000}.
Supersymmetry may play an important role in many aspects of brane world models
such as fine-tuning between bulk and brane cosmological constants and the stabilization of
the fifth dimension (BPS vacua are stable against perturbations).  Furthermore, 
supersymmetry and supergravity are critical aspects of string theory and hence
it should be expected that they will play an integral role in string theory 
realizations of brane world scenarios.

Although there was legitimate concern
that brane world models may be impossible to realize as BPS or non-BPS configurations
of a supersymmetric theory \cite{ref:kalloshet2000,ref:gibbonset2000}, recent
work has found a way to circumvent these no-go theorems (see, e.g. \cite{ref:duffet2000k}).
In \cite{ref:duffet2000k} the authors obtain the original Randall-Sundrum configuration
from type IIB supergravity.  This is achieved by considering a solution to the $D=10$
type IIB supergravity equations which has a $5$D interpretation.  Note however that
this is not fully a
$D=5$ solution as it requires the $S^5$ massive Kaluza-Klein breathing mode.
Breathing modes of sphere reductions are often useful in supporting domain walls 
\cite{ref:duffet2000k, ref:alwisetf2000, ref:liueta2000, ref:cveticetij2000}.
In this model it is possible to recover the single brane RSII model by pushing the hidden brane off to the Cauchy horizon
of AdS.  Another pleasing feature of this model is that the D3-brane configuration is dynamically stable.

Another interesting work provides a supersymmetric version of the minimal RS model in which the branes are singular 
\cite{ref:altendorferet2000}.  

Note that not all scenarios involving large extra dimensions rely
on supersymmetry, such as the ADD model described in \cite{ref:dvaliet1999a}.
The ADD scenario is not without its own troubles however as it
has light KK gravitinos which could cause drastic problems with 
nucleosynthesis and the cosmic gamma ray background \cite{ref:cline1999}.

As an increasing number of works on the supersymmetrization of the RS model become 
available we will no doubt gain a better understanding of how this configuration should be
assimilated into models of M/superstring theory and the AdS/CFT correspondence.

\section{Conclusions}\label{conclusion}

In this review we have discussed a number of intriguing approaches to string and M-theory 
cosmology.  While  
the past few years have shown a considerable increase in our understanding
of M-theory, there is still plenty of room for further research. 

Perhaps the greatest advances have come from the discovery of duality symmetries
in the M-theory moduli space, D-branes, the AdS/CFT correspondence and
the development of Matrix theory.  As demonstrated in this review we have taken
the first steps to incorporate this new knowledge into cosmology.  
M-theory provides an innovative framework in which to study 
the early Universe and to search for 
alternatives to the Standard Big-Bang and Inflationary models.
Conversely, cosmology is
essential to our study of M-theory, since couplings and masses set by the vacuum 
state of string theory must agree with those observed in our Universe.  
The amalgamation of M-theory and cosmology may reveal the answers to a 
number of tantalizing questions
and provide the tools to probe the earliest moments of creation.

\section*{Acknowledgments}\label{ack}

I wish to thank D. Dooling, A. Jevicki and especially R.
Brandenberger and D. Lowe for many useful comments and
discussions.  I would also like to thank the PIMS at the University
of British Columbia and the members of the string cosmology conference 
held there in July 2000.  In particular, I am grateful to E. Akhmedov, S. Alexander,  A. Ghosh, B. Greene, 
T. Hertog, B. Ovrut, G. Veneziano, H. Verlinde and T. Wiseman for helpful discussions.  
This work was supported in part by the Graduate
Assistance in Areas of National Need (GAANN) doctoral fellowship program.

\footnotesize

\end{document}